\documentclass[10pt,journal,compsoc,twoside]{IEEEtran}

\usepackage[ieee,algorithmic]{definition}
\usepackage[caption=false,font=footnotesize,labelfont=sf,textfont=sf]{subfig}

\newcommand{\themodel}{MICRO\xspace}

\begin{document}
\title{Latent Structure Mining with Contrastive Modality Fusion for Multimedia Recommendation}

\author{Jinghao Zhang, Yanqiao Zhu, \IEEEmembership{Student Member, IEEE,} Qiang Liu, \IEEEmembership{Member, IEEE,}\\
Mengqi Zhang, Shu Wu, \IEEEmembership{Senior Member, IEEE,} and Liang Wang, \IEEEmembership{Fellow, IEEE}%
\IEEEcompsocitemizethanks{ \IEEEcompsocthanksitem This paper is an extended version of LATTICE \cite{zhang_lattice}, which has been published at the Proceedings of the 29th ACM International Conference on Multimedia. 
\IEEEcompsocthanksitem All authors are with the Center for Research on Intelligent Perception and Computing, Institute of Automation, Chinese Academy of Scienceszhenz and are also with the School of Artificial Intelligence, University of Chinese Academy of Sciences.
\IEEEcompsocthanksitem Corresponding author: Qiang Liu (qiang.liu@nlpr.ia.ac.cn). 
}\\
\thanks{Manuscript received April 19, 2005; revised August 26, 2015.}}

\markboth{In Submission to IEEE Transactions on Knowledge and Data Engineering}%
{Zhang \MakeLowercase{\textit{et al.}}: Latent Structure Mining with Contrastive Modality Fusion for Multimedia Recommendation}

\IEEEtitleabstractindextext{%
\begin{abstract}
Multimedia content is of predominance in the modern Web era. Recent years have witnessed growing research interests in multimedia recommendation, which aims to predict whether a user will interact with an item with multimodal contents. Most previous studies focus on modeling user-item interactions with multimodal features included as side information. However, this scheme is not well-designed for multimedia recommendation. Firstly, only \emph{collaborative} item-item relationships are implicitly modeled through high-order item-user-item co-occurrences. Considering that items are associated with rich contents in multiple modalities, we argue that the latent \emph{semantic} item-item structures underlying these multimodal contents could be beneficial for learning better item representations and assist the recommender models to comprehensively discover candidate items.  Secondly, previous studies disregard the fine-grained multimodal fusion. Although having access to multiple modalities might allow us to capture rich information, we argue that the simple fusion by linear combination or concatenation in previous work is insufficient to fully understand content information of items and item relationships.To this end, we propose a latent structure \underline{MI}ning with \underline{C}ont\underline{R}astive m\underline{O}dality fusion method, which we term \themodel for brevity. To be specific, in the proposed \themodel model, we devise a novel modality-aware structure learning module, which learns item-item relationships for each modality. Based on the learned modality-aware latent item relationships, we perform graph convolutions which explicitly inject item affinities to modality-aware item representations. Additionally, we design a novel multi-modal contrastive framework to facilitate fine-grained multimodal fusion by forcing the modality-aware representation and multimodal fused representation to be close. Finally, these enriched item representations can be plugged into existing collaborative filtering methods to make more accurate recommendations. Extensive experiments on three real-world datasets demonstrate the superiority of our method over state-of-the-art multimedia recommendation methods and ablation studies validate the efficacy of mining latent item-item relationships and the contrastive multimodal fusion framework. 
\end{abstract}

}

\maketitle

\IEEEdisplaynontitleabstractindextext

\newcommand{\inlinegraphics}[1]{(\raisebox{-.1\height}{\includegraphics[width=0.9em]{#1}})}

\IEEEraisesectionheading{\section{Introduction}\label{sec:introduction}}

\IEEEPARstart{W}{ith} the rapid development of Internet, information overload has become an increasingly crucial challenge. Personalized recommender systems act as an indispensable tool to help users find their preferred information from massive irrelevant contents. Nowadays, users are easily accessible to large amounts of online information represented in multiple modalities, including images, texts, videos, etc. For example, the \emph{visual appearance} and \emph{textual descriptions} play important roles when users selecting products online; the \emph{visual cover} and \emph{textual tags} allow users to find interesting items from a large amount of instant videos. Recent years have witnessed growing research interests in multimedia recommendation, which aims to predict whether a user will interact with an item with multimodal contents. It has been successfully applied to many online applications, such as e-commerce, instant video platforms and social media platforms.

\begin{figure}[t]
	\centering
	\includegraphics[width=\linewidth]{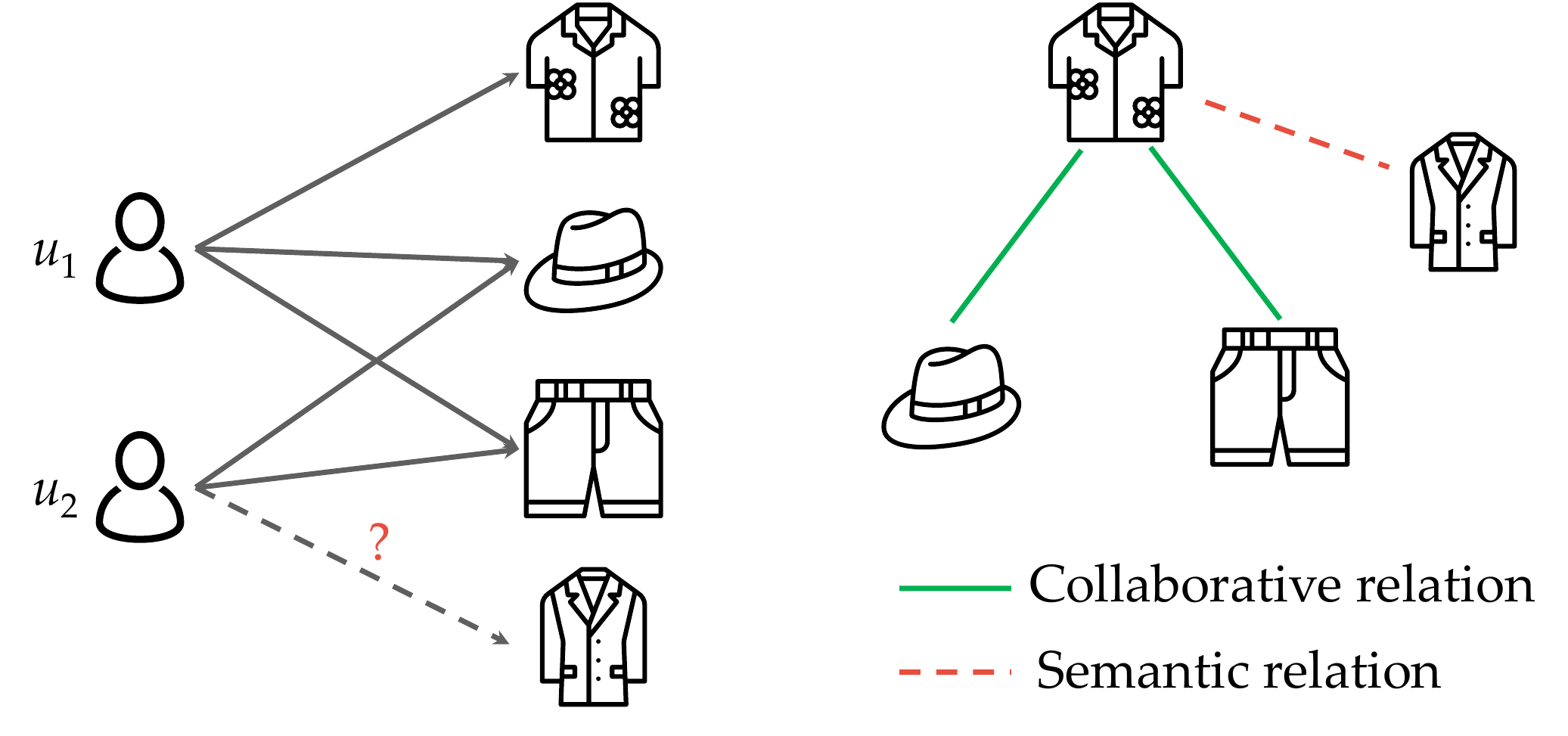}
	\caption{A toy example of recommendation with two types of item relations. In this paper, we argue that {\color[rgb]{0.89803922,0.30980392,0.26666667} semantic structures} mined from multimodal features are helpful for comprehensively discovering candidate items supplementary to {\color[rgb]{0.09803922,0.68627451,0.32941176} collaborative signals} in traditional work (Best viewed in color).}
	\label{fig:toy-example}
\end{figure}

Collaborative Filtering (CF), as one of the most prevalent techniques in personalized recommendation, have been widely studied. Focusing on exploiting abundant user-item interactions, CF methods group users according to their historical interactions, by encoding users and items into low-dimensional dense vectors and making recommendations based on these embeddings \cite{Aggarwal:2016dl,He:2017jw,Su:2009cl}. Following traditional CF framework, early work on multimedia recommendation like VBPR \cite{He:2016ww}, DeepStyle \cite{Liu:2017ij}, and ACF \cite{Chen:2017jj} incorporates multimodal features as side information in addition to the learned dense vectors of items, so as to group users based on both historical interactions and item contents. \citet{park2017also} propose to explicitly capture the information hidden in also-viewed products, i.e., a list of products that have also been viewed by users who have viewed a target product.  The also-viewed relationship can be regarded as a special kind of co-interacted relationship. \citet{Lee2017LargeScaleCV} propose to pull the content vectors of co-watched items to be closer, which exploits co-interacted item-item relationships through item-user-item occurrences.

Inspired by the recent surge of graph neural networks \cite{Kipf:2017tc,Velickovic:2018we}, \citet{Wang:2019er} propose to model user-item relationships as bipartite graphs. The first-order connectivities in user-item graphs indicate the interaction history. And the second-order connectivities reveal collaborative relations that similar users (or items) who have co-interacted with the same items (or users). These graph-based recommender systems \cite{Wang:2019er,Wu:2019ke,He:2020gd} inject high-order connectivities into the embedding process to learn better representations and achieve great success. Recently, many attempts have been made to integrate multimodal contents into graph-based recommendation systems. MMGCN \cite{Wei:2019hn} constructs modality-specific user-item interaction graphs to model user preferences specific to each modality. Following MMGCN, GRCN \cite{Wei:2020ko} utilizes multimodal features to refine user-item interaction graphs by identifying false-positive feedbacks and prunes the corresponding noisy edges. HUIGN \cite{wei2021hierarchical} constructs a co-interacted item graph, where the edge corresponds to the item pair consumed by the same users. By conducting hierarchical GNNs on the co-interacted item graph, HUIGN can mine users’ intents at different levels.

Despite their effectiveness, previous attempts suffer from two limitations. Firstly, existing work fails to comprehensively model item-item relationships, which have been proved to be important in recommender systems \cite{Sarwar:2001kx}. Specifically, only \textbf{collaborative} relations are considered through high-order item-user-item co-occurrences \cite{park2017also, Lee2017LargeScaleCV, wei2021hierarchical}. However, \textbf{semantic} relations which reflect the content information of items, are not explicitly modeled. Taking Figure \ref{fig:toy-example} as an example, existing methods will recommend the shirt \inlinegraphics{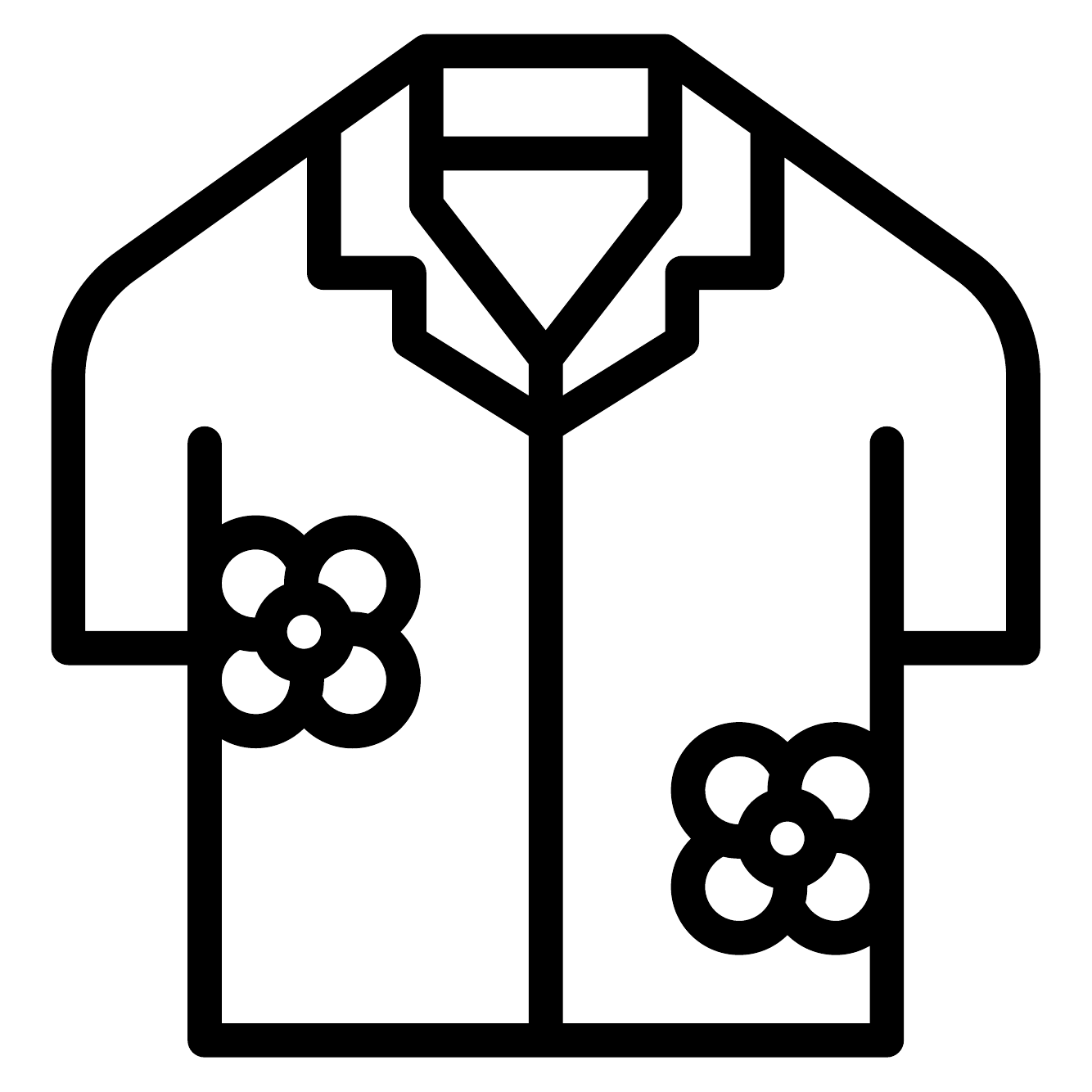} for \(u_2\) according to collaborative relations, since shirts \inlinegraphics{figures/icons/hawaiian-shirt.pdf}, hats \inlinegraphics{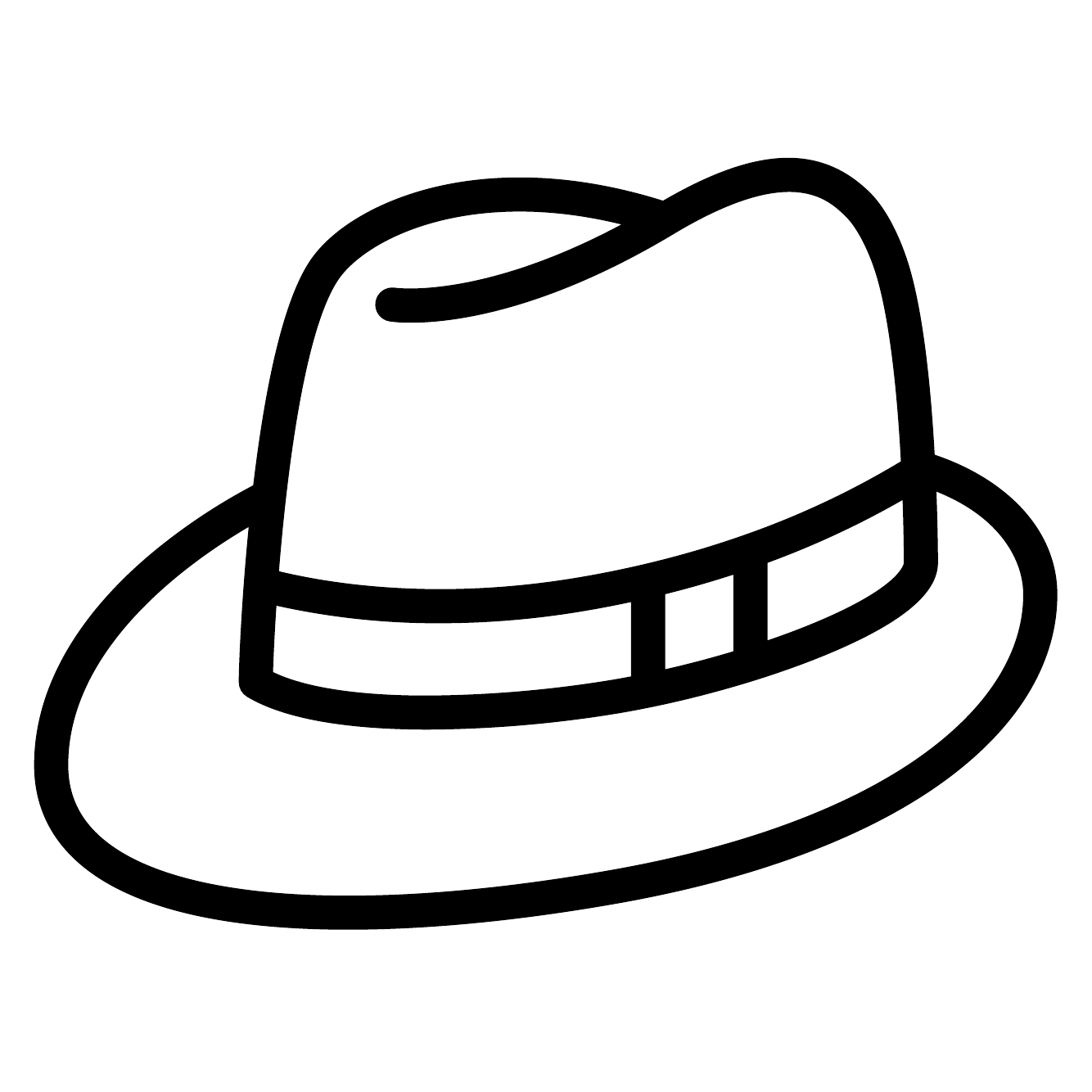}, and pants \inlinegraphics{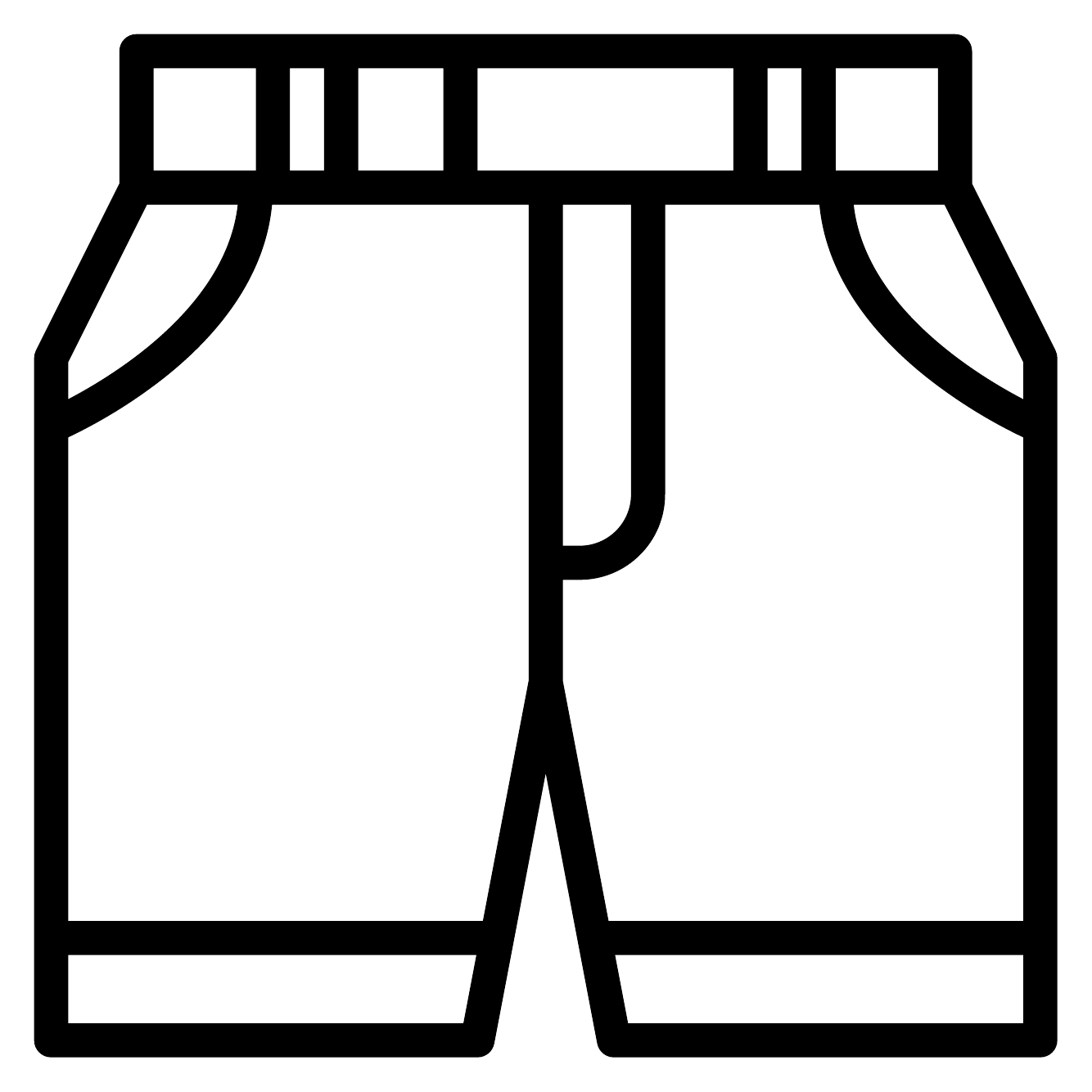} all interacted with \(u_1\). However, previous work may not be able to recommend coats \inlinegraphics{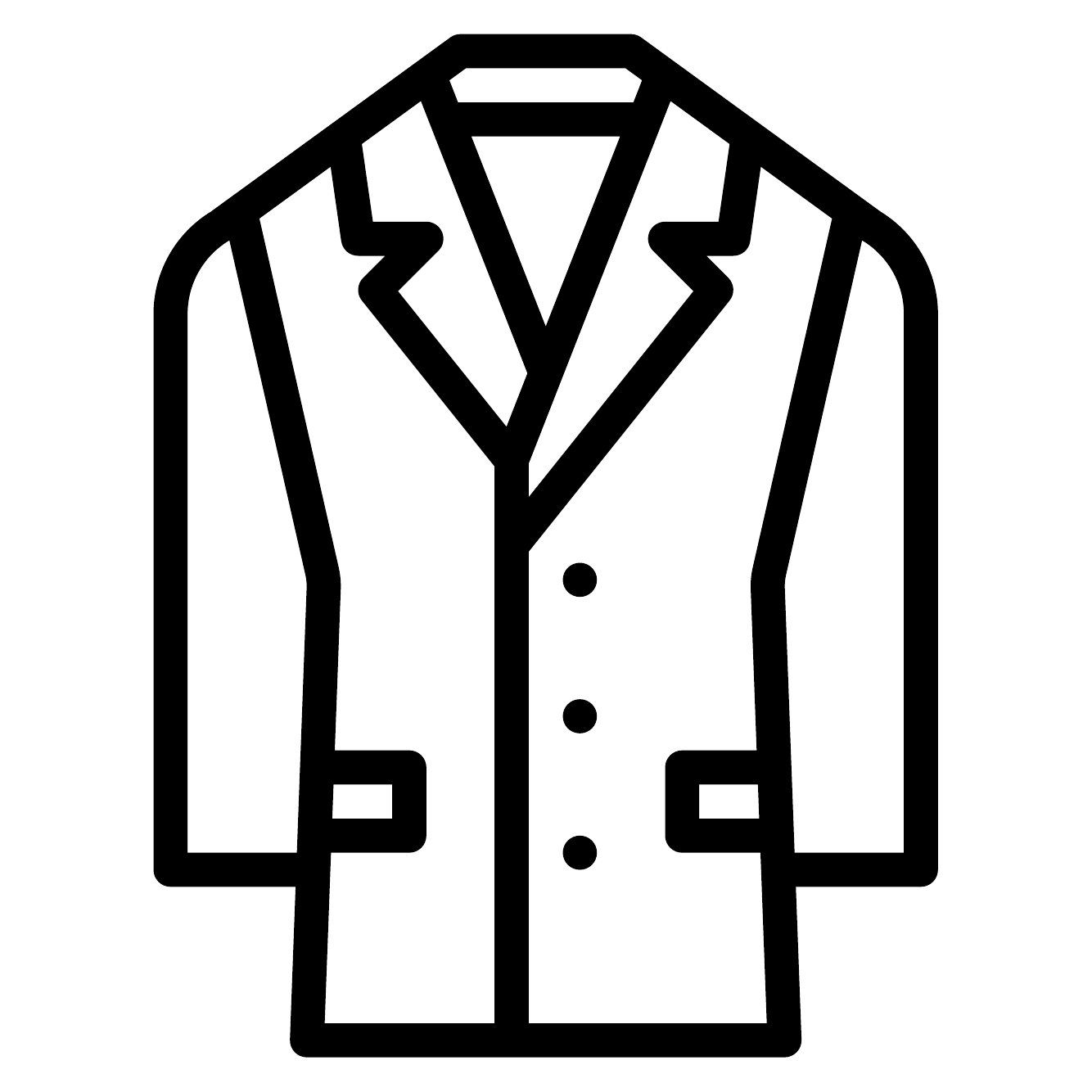} to \(u_2\), which are semantically (visually in this example) similar to shirts. Considering that items are associated with rich multimodal content features in multimedia recommendation, there exist a wealth of semantic relations underlying multimodal contents, which would assist the recommender models to comprehensively discover candidate items. 

Secondly, previous attempts disregard the fine-grained multimodal fusion. Early work \cite{He:2016ww, Liu:2017ij, he2016sherlock} only focuses on unimodal information; other work on multimedia recommendation \cite{Wei:2019hn, Wei:2020ko} conducts coarse-grained multimodal fusion by simple linear combination or concatenation, the inductive bias behind which is that all items share the same fusion mechanism (e.g., the same combination weights). On the contrary, we conduct fine-grained multimodal fusion at the item level, permitting the model to utilize the most relevant parts of different items in a flexible manner and therefore learn better item representations. Based on the hypothesis that a powerful representation is the one that models modality-invariant factors \cite{smith2005development}, we propose to conduct fine-grained multimodal fusion by capturing the \emph{shared item relationships} from multiple modalities. There is significant evidence in the cognitive and neuroscience areas that such modality-invariant representations are encoded by the brain \cite{smith2005development, hohwy2013predictive} and has exhibited remarkable benefits in some multimodal tasks \cite{tian2020contrastive, alwassel_2020_xdc, morgado2021audio}. 

In this paper, we propose a novel method to mine latent semantic item-item relationships underlying multimodal features of items, and conduct fine-grained multimodal fusion based on the learned structures to inject shared item-item relationships from multiple modalities into the item representations. As shown in Figure \ref{fig:model}, the proposed \themodel consists of four key components. Firstly, we develop a novel modality-aware structure learning layer, which learns modality-aware item structures from content features of each modality. Secondly, we perform graph convolutions on the learned modality-aware latent graphs to explicitly consider item relationships of each modality individually. Thirdly, we devise a novel multimodal contrastive framework and construct self-supervision signals by maximizing the agreement between item representations under individual modalities and the multimodal fused representations, and thus the fused multimodal representations can adaptively capture item-item relationships \emph{shared between multiple modalities} in a self-supervised manner. Finally, the resulting enhanced item representations are infused with item relationships in multiple modalities, which will be added into the output item embeddings of CF models to make recommendations.

Our work enjoys two additional benefits. Firstly, \themodel can alleviate the cold-start problem. Previous graph-based multimedia recommendation methods face cold-start problems where long-tailed items are only interacted with few users or even never interacted with users. Since previous methods utilize multimodal content features based on user-item interaction graph, those long-tailed items will become isolated nodes in graph, which will invalidate their usage of multimodal information. Our work, on the contrary, can alleviate the cold-start problem in two ways: firstly, we mine latent item-item structures and the long-tailed items will get similar user feedbacks from their learned neighbors; secondly, the multimodal contrastive framework serves as a self-supervised auxiliary task. The external self-supervision signals are introduced to learn better item representations involved with relation information, which would further alleviate the cold-start problem. 

Secondly, \themodel can serve as a flexible play-and-plug module. Unlike previous attempts which utilize multimodal features based on dedicated user-item aggregation strategies, \themodel separates the usage of multimodal features with the usage of user-item interactions and is agnostic to downstream CF methods.

In summary, the main contribution of this work is threefold.
\begin{itemize}
	\item We highlight the importance of explicitly exploiting item relationships and considering fine-grained multimodal fusion in multimedia recommendation.
	\item We propose a novel method to mine latent item relations and conduct fine-grained multimodal fusion based on the mined structures.
	\item We perform extensive experiments on three public datasets. Notably, our method outperforms the state-of-the-art methods by 20\% on average in terms of different metrics, validating the effectiveness of our proposed model.
\end{itemize}
To foster reproducible research, our code is made publicly available at \url{https://github.com/CRIPAC-DIG/MICRO}.

\section{Preliminaries}
In this section, we first formulate the multimedia recommendation problem. Secondly, to make our motivation more convincing, we use two simple and intuitive experiments, from item perspective and user perspective, to show that users tend to buy semantically similar items. That is, semantic item-item relationships are helpful for comprehensively discovering candidate items.

\subsection{Notations}
Let $\mathcal{U}$, $\mathcal{I} (|\mathcal{I}| = N)$ denote the set of users and items, respectively.  Each user $u \in \mathcal{U}$ is associated with a set of items $\mathcal{I}^u$ with positive feedbacks which indicate the preference score $y_{ui}=1$ for $i \in \mathcal{I}^u$. $\bm x_u, \bm x_i \in \mathbb{R}^{d}$ is the input ID embedding of $u$ and $i$, respectively, where $d$ is the embedding dimension. Besides user-item interactions, multimodal features are offered as content information of items. We denote the modality features of item $i$ as $\bm{e}_i^m \in \mathbb{R}^{d_m}$, where $d_m$ denotes the dimension of the features, $m \in \mathcal{M}$ is the modality, and $\mathcal{M}$ is the set of modalities.
The purpose of multimedia recommendation is to accurately predict users' preferences by ranking items for each user according to predicted preference scores $\hat{y}_{u i}$.
In this paper, we consider visual and textual modalities denoted by $\mathcal{M} = \{ \text{v}, \text{t} \}$. Please kindly note that our method is not fixed to the two modalities and multiple modalities can be involved.

\subsection{Pilot Studies}
Firstly, from item perspective, we conduct an experiment to show that co-interacted items are much more semantically similar. We compute the cosine similarity between all items as baseline and compute the similarity between co-interacted items. The averages are summarized in Table 1. We can observe that co-interacted items are much more similar. This indicates that the items bought by the same user are much more similar. That is, users tend to buy semantically similar items.

\begin{table}
\centering
\caption{Average semantic similarity of all items and co-interacted items.}
\label{tab:avgsim}
\begin{tabular}{@{}cccc@{}}
\toprule
Dataset & Modality  & All Items & Co-interacted Items  \\ \midrule
\multirow{2}{*}{Clothing} 
& Visual               &  0.2239         &   0.3958      \\
& Textual                &  0.4206         &   0.5830      \\    \midrule
\multirow{2}{*}{Sports} 
& Visual               &  0.2184         &   0.3547   \\
& Textual                &  0.3895         &   0.5423    \\    \midrule
\multirow{2}{*}{Baby} 
& Visual               &  0.2240         &   0.3534    \\
& Textual                &  0.4413         &   0.5405     \\    \bottomrule
\end{tabular}
\end{table}

Secondly, from user perspective, we count the proportion of users buying semantically similar items.  We intuitively define $i_1$ and $i_2$ are semantically similar if $i_1$ is among the $k$ items most similar to $i_2$, or $i_2$ is among the $k$ items most similar to $i_1$, where a smaller $k$ means a smaller range. Table \ref{tab:proportion} reports the proportion of users buying semantically similar items with respect to different $k$. We can observe that even with a small $k$, the majority of users tend to buy semantically similar items.

\begin{table}
\centering
\caption{The proportion ($\%$) of users buying semantically similar items with respect to different $k$.}
\label{tab:proportion}
\begin{tabular}{@{}cccccc@{}}
\toprule
Dataset & Modality  & $k=5$ & $k=10$  & $k=15$ & $k=20$ \\ \midrule
\multirow{2}{*}{Clothing} 
& Visual               &  46.88         &   51.34   &   54.22   &   56.33      \\
& Textual                &  54.90         &   60.21   &   63.41   &   65.57      \\    \midrule
\multirow{2}{*}{Sports} 
& Visual               &  42.18         &   45.58   &   47.71   &   49.39   \\
& Textual                &  53.56         &   58.48   &   61.64  &   63.90    \\    \midrule
\multirow{2}{*}{Baby} 
& Visual               &  44.37         &   48.17   &   50.87   &   53.36    \\
& Textual                &  55.25         &   59.58   &   62.45   &   64.89     \\    \bottomrule
\end{tabular}
\end{table}
\section{The Proposed Method}
\begin{figure*}
	\centering
    \includegraphics[width=0.9\linewidth]{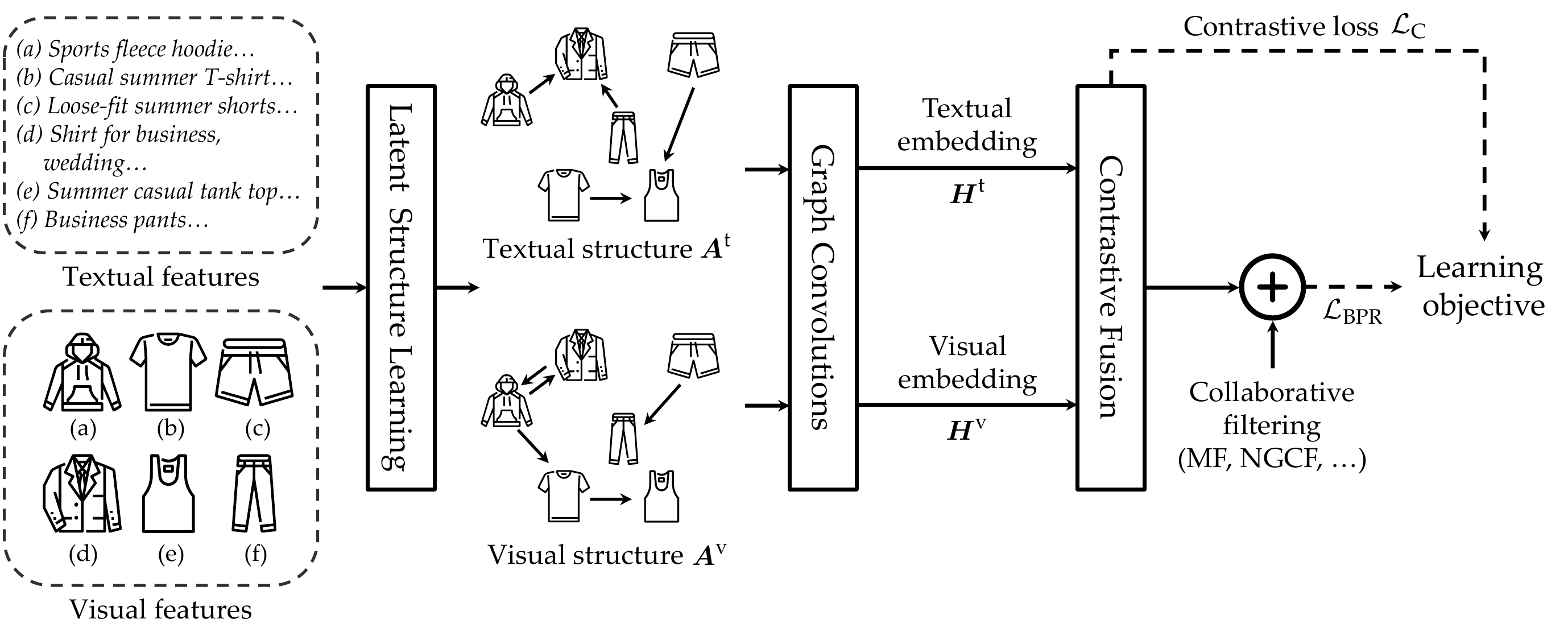}
    \caption{The overall framework of our proposed \themodel model.  Firstly, we develop a novel modality-aware structure learning layer to mine the modality-aware latent item-item semantic relationships from multimodal features. Secondly, we employ graph convolutions on the learned modality-aware graphs to explicitly model item relationships of each modality individually. Thirdly, we devise a novel contrastive multimodal fusion framework to force the fused multimodal representations to adaptively capture item relationships shared between multiple modalities in a self-supervised manner. Finally, the resulting item representations are infused with item relationships in multiple modalities, which will be added into the output item embeddings of CF models to make recommendations. The contrastive loss and recommendation (BPR) loss will be optimized together.}
    \label{fig:model}
\end{figure*}
In this section, we introduce our model in detail. As illustrated in Figure \ref{fig:model}, there are four main components in our proposed framework:
(1) a modality-aware graph structure learning layer that learns item graph structures from content features of each modality,
(2) graph convolutional layers that learn the modality-aware item embeddings by injecting item-item affinities based on the learned graph structures,
(3) an attentive multimodal fusion framework with contrastive auxiliary task to promote fine-grained multimodal fusion, 
and (4) downstream CF methods. 

\subsection{Modality-aware Latent Structure Mining}
Multimodal features provide rich and meaningful content information of items, while existing methods only utilize multimodal features as side information for each item, ignoring the important \emph{semantic} relationships of items underlying features. In this section, we introduce how to discover the underlying latent graph structure of item graphs in order to learn better item representations.

To be specific, we first construct initial $k$-Nearest-Neighbor ($k$NN) modality-aware item graphs $\widetilde{\bm{S}}^m$ by utilizing raw multimodal features. After that, we learn the latent graph structures $\widetilde{\bm{A}}^m$ from transformed multimodal features. Finally, we combine the learned structures with the initial structures by a skip connection.

\subsubsection{Constructing Initial Modality-aware Graphs}
We first construct initial $k$NN modality-aware graph $\bm{S}^m$ by using raw features for each modality $m$. Based on the hypothesis that similar items are more likely to interact than dissimilar items \cite{McPherson:2003dp}, we quantify the \emph{semantic} relationship between two items by their similarity.
Common options for node similarity measurement include cosine similarity \cite{Wang:2020bs}, kernel-based functions \cite{Li:2018wu}, and attention mechanisms \cite{Chen:2020wu}. Our method is agnostic to similarity measurements, and we opt to the simple and parameter-free cosine similarity in this paper. The similarity matrix $\bm{S}^m \in \mathbb{R}^{N \times N}$ is computed by
\begin{equation}
    \bm{S}_{i j}^m = \frac{{(\bm{e}_i^m)}^\top \bm{e}_j^m}{ \| \bm{e}_i^m \| \| \bm{e}_j^m \|}.
    \label{eq:sim}
\end{equation}
Typically, the graph adjacency matrix is supposed to be non-negative but $\bm{S}_{i j}$ ranges between $[-1, 1]$. Thus, we suppress its negative entries to zeros.
Moreover, common graph structures are much sparser other than a fully-connected graph, which is computationally demanding and might introduce noisy, unimportant edges \cite{Chen:2020wu}. We conduct $k$NN sparsification \cite{Chen:2009wc} on the dense graph: for each item $i$, we only keep edges with the top-\(k\) confidence scores:
\begin{equation}
    \widehat{\bm{S}}^m_{i j}=
    \begin{cases}
    \bm{S}^m_{i j}, \enspace & {\bm{S}}^m_{i j} \in \operatorname{top-}k({\bm{S}}^m_{i,:}), \\
    0, \enspace & \text{otherwise},
    \end{cases}
    \label{eq:topk}
\end{equation}
where ${\bm{S}}^m_{i,:}$ denotes the $i$-row of ${\bm{S}}$, and $\widehat{\bm{S}}^m$ is the resulting sparsified, directed graph adjacency matrix. To alleviate the exploding or vanishing gradient problem \cite{Kipf:2017tc}, we normalize the adjacency matrix as:
\begin{equation}
    \widetilde{\bm{S}}^m = ({\bm{D}^m})^{-\frac{1}{2}} \widehat{\bm{S}}^m ({\bm{D}^m})^{-\frac{1}{2}},
    \label{eq:norm}
\end{equation}
where $\bm{D}^m \in \mathbb{R}^{N \times N}$ is the diagonal degree matrix of $\widehat{\bm{S}}^m$ and $\bm{D}_{ii}^m = \sum_{j}\widehat{\bm{S}}^m_{i j}$.

\subsubsection{Learning Latent Modality-aware Graphs}
Although we have obtained the modality-aware initial graph structures $\widetilde{\bm{S}}^m$ by utilizing raw multimodal features, they may not be ideal for the recommendation task. This is because the raw multimodal features are often noisy or even incomplete due to the inevitably error-prone data measurement or collection. To this end, we propose to dynamically learn the graph structures by the  transformed multimodal features and combine the learned structures with initial ones.

Firstly, we transform raw modality features into high-level features $\widetilde{\bm{e}}_{i}^m$:
\begin{equation}
    \widetilde{\bm{e}}_{i}^m = \bm{W}_m {\bm{e}}_{i}^m + \bm{b}_m,
\end{equation}
where ${\bm{W}}_m \in \mathbb{R}^{d \times d_m}$ and ${\bm{b}}_m \in \mathbb{R}^{d}$ denote the trainable transformation matrix and the bias vector, respectively. We dynamically infer the graph structures utilizing $\widetilde{\bm{e}}_{i}^m$, repeat the graph learning process described in Eqs. (\ref{eq:sim}, \ref{eq:topk}, \ref{eq:norm}) and obtain the adjacency matrix $\widetilde{\bm{A}}^m$.

Although the initial graph could be noisy, it still carries rich and useful information regarding item graph structures. Also, drastic change of adjacency matrix will lead to unstable training. To keep rich information of initial item graph and stabilize the training process, we add a skip connection that combines the learned graph with the initial graph:
\begin{equation}
    \bm{A}^m =  \lambda \widetilde{\bm{S}}^m + (1 - \lambda) \widetilde{\bm{A}}^m \ ,
\end{equation}
where $\lambda \in (0, 1)$ is the coefficient of skip connection that controls the amount of information from the initial structure. The obtained $\bm{A}^m$ is the final graph adjacency matrix representing latent structures for modality \(m\).

It is worth mentioning that both $\widetilde{\bm{S}}^m$ and $\widetilde{\bm{A}}^m$ are sparsified and normalized matrices, thus the final adjacency matrix $\bm{A}^m$ is also sparsified and normalized, which is computationally efficient and stabilizes gradients.

\subsection{Item Affinities Learning with Graph Convolutions}
\label{sec:graph-conv}
After obtaining the modality-aware latent structures, we perform graph convolution operations to learn better item representations by injecting item-item affinities into the embedding process. Graph convolutions can be treated as message propagation and aggregation. Through propagating the item representations from its neighbors, one item can aggregate information within the first-order neighborhood. Furthermore, by stacking multiple graph convolutional layers, the high-order item-item relationships can be captured.

Following \citet{Wu:2019vz} and \citet{He:2020gd}, we employ simple message propagation and aggregation without feature transformation and non-linear activations which is effective and computationally efficient. In the $l$-th layer, the message passing and aggregation could be formulated as:

\begin{equation}
    \bm{H}^{m}_{(l)} =  \bm{A}^m \bm{H}^{m}_{(l-1)},
\end{equation}
where $\bm{H}^{m}_{(l)} \in \mathbb{R}^{N \times d}$ is the $l$-th layer item embedding matrix of modality $m$, the $i$-th row of which denotes the embedding vector of item $i$. For all modalities $m \in \mathcal{M}$, we use the same item ID embedding matrix to initialize the input embedding matrix $\bm{H}^{m}_{(0)}$. We utilize ID embedding vector of items as input representations rather than multimodal features, since we employ graph convolutions to directly capture item-item affinities and multimodal features are used to bridge semantic relationships. After stacking $L$ layers, $\bm{H}^{m}_{(L)}$ encodes the high-order item-item relationships of modality $m$.

\subsection{Multimodal Fusion with Contrastive Auxiliary Task}
\label{sec:fusion}
Multiple modalities convey comprehensive information \cite{baltruvsaitis2018multimodal}. Item relations shared between modalities are important to learn better item representations based on the hypothesis that a powerful representation is one that models modality-invariant factors \cite{smith2005development, tian2020contrastive}. To this end, we first utilize an attention mechanism to fuse item embeddings $\bm{H}^{m}_{(L)}$ of different modalities, and then devise a self-supervised contrastive loss to promote multimodal fusion.  
\subsubsection{Aggregating Multiple Modalities}
We omit the subscript $(L)$ and use $\bm{h}^m_i$ to denote the $i$-th row of $\bm{H}^{m}_{(L)}$, which is the output embedding of graph convolutions corresponding to item $i$. The importance of each modality corresponding to item $i$ can be formulated as follows:
\begin{equation}
w_{i}^{m}= \boldsymbol{q}^{\top} \tanh \left(\boldsymbol{W} \boldsymbol{h}_{i}^{m}+\boldsymbol{b}\right),
\end{equation}
where $\bm{q} \in \mathbb{R}^{d}$ denotes attention vector and $\bm{W} \in \mathbb{R}^{d \times d}, \bm{b} \in \mathbb{R}^{d}$ denote the weight matrix and bias vector, respectively. These parameters are shared for all modalities. After obtaining the importance of different modalities, we normalize them to get the weight coefficients:
\begin{equation}
    \alpha_{i}^{m}=\frac{\exp \left(w_{i}^{m}\right)}{\sum_{m=1}^{|\mathcal{M}|} \exp \left(w_{i}^{m}\right)},
\end{equation}
Then, the multimodal fused embedding of item $i$ can be represented as:
\begin{equation}
    \bm{h}_{i}=\sum_{m=1}^{|\mathcal{M}|} \alpha_{i}^{m} \boldsymbol{h}_{i}^{m}.
\end{equation}

\subsubsection{Contrastive Auxiliary Task}
\label{sec:contrast}
After obtaining the multimodal fused item embeddings, we devise a novel self-supervised auxiliary task to further force the fused item embeddings to adaptively distill the shared information from multiple modalities. Existing contrastive learning frameworks \cite{Chen:2020wj} seek to maximize the agreement among differently augmented views of the same data examples, which has been proven to be effective in multi-view representation learning \cite{tian2020contrastive, stojnic2021self} and multimodal tasks \cite{liu2021contrastive, whitehead2021separating}. In this work, since multiple modality-aware graphs are involved, we propose to construct self-supervision signals by \textit{maximizing the agreement between item representations under individual modalities and the fused multimodal representations}. In this way, the fused multimodal representations can adaptively capture item-item relationships \emph{shared between multiple modalities} in a self-supervised manner. The resulting contrastive loss can be mathematically expressed as:
\begin{equation}
    \small
    \mathcal{L}_{\text{C}}=- \frac{1}{|\mathcal{I}|} \sum_{i \in \mathcal{I}}\left[\frac{1}{|\mathcal{M}|} \sum_{m \in \mathcal{M}} \frac{1}{2}\left(I\left(\boldsymbol{h}_{i}^{m} , \boldsymbol{h}_{i}\right)+I\left(\boldsymbol{h}_{i} , \boldsymbol{h}_{i}^{m}\right)\right)\right], 
\end{equation}
where $I(\cdot , \cdot)$ denotes the mutual information which quantifies the agreement between two representations, which is implemented by the InfoNCE estimator \cite{Chen:2020wj}. Specifically, for $I(\boldsymbol{h}_{i}^{m}, \boldsymbol{h}_{i}) $, we set $(\boldsymbol{h}_{i}^{m}, \boldsymbol{h}_{i})$ as positive samples, while all other item embeddings in an individual modality $(\boldsymbol{h}_{i}^{m}, \boldsymbol{h}_{j}^m)$ and the fused multimodal embeddings $(\boldsymbol{h}_{i}^{m}, \boldsymbol{h}_{j})$ are considered as negatives:
\begin{equation}
\small
\begin{aligned}
I(&\boldsymbol{h}_{i}^{m}, \boldsymbol{h}_{i}) = \\ 
& \log \frac{e^{\theta\left(\boldsymbol{h}_{i}^{m}, \boldsymbol{h}_{i}\right) / \tau}}{e^{\theta\left(\boldsymbol{h}_{i}^{m}, \boldsymbol{h}_{i}\right) / \tau}+\sum_{j \neq i}\left(e^{\theta\left(\boldsymbol{h}_{i}^{m}, \boldsymbol{h}_{j}\right) / \tau}+e^{\theta\left(\boldsymbol{h}_{i}^{m}, \boldsymbol{h}_{j}^{m}\right) / \tau}\right)},
\end{aligned}
\end{equation}
where $\tau \in \mathbb{R}$ is a temperature parameter and $\theta(\cdot, \cdot)$ is the critic function which is implemented by a simple cosine similarity. Similarly,  $I(\boldsymbol{h}_{i}, \boldsymbol{h}_{i}^{m})$ can be formulated as:
\begin{equation}
\small
\begin{aligned}
I(&\boldsymbol{h}_{i}, \boldsymbol{h}_{i}^{m}) = \\ 
& \log \frac{e^{\theta\left(\boldsymbol{h}_{i}^{m}, \boldsymbol{h}_{i}\right) / \tau}}{e^{\theta\left(\boldsymbol{h}_{i}^{m}, \boldsymbol{h}_{i}\right) / \tau}+\sum_{j \neq i}\left(e^{\theta\left(\boldsymbol{h}_{i}^{m}, \boldsymbol{h}_{j}\right) / \tau}+e^{\theta\left(h_{i}, h_{j}\right) / \tau}\right)}
\end{aligned}
\end{equation}

The proposed objective also conceptually relates to contrastive knowledge distillation \cite{tian2019contrastive}, where several teacher models (representations under different individual modalities) and one student model (the fused representations) are employed. By forcing the embeddings between several teachers and a student to be the same, these fused representations adaptively collect information from all modality-aware item relations. Additionally, the multimodal contrastive framework serves as a self-supervised auxiliary task, where the external self-supervision signals are introduced to learn better item representations involved with relation information from multiple modalities, which would further alleviate the cold-start problem.

\subsection{Incorporating with Collaborative Filtering Methods}
\label{CF}
Unlike previous attempts which utilize multimodal features based on sophisticated user-item aggregation strategies, \themodel separates the usage of multimodal features with the usage of user-item interactions and is agnostic to downstream CF methods. Specifically, we learn item representations from mined item relations and then combine them with downstream CF methods that model user-item interactions. It is flexible and could be served as a play-and-plug module for any CF methods. 

We denote the output user and item embeddings from CF methods as $\widetilde{\bm{x}}_u, \widetilde{\bm{x}}_i \in \mathbb{R}^d$ and simply enhance item embeddings by adding normalized multimodal fused item embeddings $\bm{h}_{i}$:
\begin{equation}
    \widehat{\bm{x}}_{i} = \widetilde{\bm{x}}_i + \frac{\bm{h}_{i}}{\|\bm{h}_{i}\|_2}.
    \label{eq:cf}
\end{equation}
We then compute the user-item preference score by taking inner product of user embeddings and enhanced item embeddings:
\begin{equation}
    \hat{y}_{u i} = \widetilde{\bm{x}}_u^\top \widehat{\bm{x}}_{i}.
\end{equation}

Additionally, the play-and-plug paradigm separates the usage of multimodal features with user-item interactions, thus alleviating the cold-start problem, where the long-tailed items are only interacted with few users or even never interacted with users. We learn latent structures for items and the tailed items will get similar feedbacks from relevant neighbors through neighborhood aggregation. 

\subsection{Optimization}
We adopt the Bayesian Personalized Ranking (BPR) loss \cite{Rendle:2009wp} to compute the pair-wise ranking, which encourages the prediction of an observed entry to be higher than its unobserved counterparts:
\begin{equation}
    \label{eq:bpr}
    \mathcal{L}_{\text{BPR}}=-\sum_{u \in \mathcal{U}} \sum_{i \in \mathcal{I}_{u}} \sum_{j \notin \mathcal{I}_{u}} \ln \sigma\left(\hat{y}_{u i}-\hat{y}_{u j}\right),
\end{equation}
where $\mathcal{I}^{u}$ indicates the observed items associated with user $u$ and $(u, i, j)$ denotes the pairwise training triples where $i \in \mathcal{I}^u$ is the positive item and $j \notin \mathcal{I}^u$ is the negative item sampled from unobserved interactions. $\sigma(\cdot)$ is the sigmoid function.

The overall loss function can be formulated as:
\begin{equation}
    \mathcal{L}=\mathcal{L}_{\text{BPR}} + \beta \mathcal{L}_{\text{C}},
\end{equation}
where $\beta \in \mathbb{R}$ is a hyper-parameter to control the effect of the contrastive auxiliary task.

\section{Experiments}
In this section, we conduct experiments on three widely used real-world datasets to answer the following research questions:
\begin{itemize}
	\item \textbf{RQ1:} How does our model perform compared with the state-of-the-art multimedia recommendation methods and other CF methods in both warm-start and cold-start settings?
	\item \textbf{RQ2:} How do the structure mining and contrastive learning modules contribute to the model performance?
	\item \textbf{RQ3:} How sensitive is our model under the perturbation of several key hyper-parameters?
	\item \textbf{RQ4:} What is the contribution of each modality?
\end{itemize}

\subsection{Experimental Settings}
\subsubsection{Datasets}

\begin{table}[t]
\centering
\begin{threeparttable}[t]
	\caption{Statistics of the datasets}
	\begin{tabular}{ccccc}
	\toprule
	Dataset\tnotex{tn:link} & \# Users & \# Items & \# Interactions & Density \\
	\midrule
	Clothing & 39,387   & 23,033   & 237,488         & 0.00026 \\
	Sports   & 35,598   & 18,357   & 256,308         & 0.00039 \\
	Baby     & 19,445   & 7,050    & 139,110         & 0.00101 \\
	\bottomrule
	\end{tabular}
	\label{tab:dataset}
	\begin{tablenotes}[flushleft]
	\footnotesize{
		\item[1]\label{tn:link} Datasets can be accessed at \url{http://jmcauley.ucsd.edu/data/amazon/links.html}.
	}
	\end{tablenotes}
\end{threeparttable}
\end{table}

\begin{table*}
	\centering
	\caption{Performance comparison of our \themodel with different baselines in terms of Recall@20 (R@20), Precision@20 (P@20), and NDCG@20. The best performance is highlighted \textbf{in bold} and the second is highlighted by \underline{underlines}. $\Delta$Improvement indicates the relative improvement of \themodel compared \underline{to the best baseline} in percentage. All improvements are significant with \(p\)-value \(\leq 0.05\).}
	\label{tab:experiments}
	\begin{tabular}{cccccccccc}
	\toprule
	\multirow{2.5}{*}{Model} & \multicolumn{3}{c}{Clothing}       & \multicolumn{3}{c}{Sports}         & \multicolumn{3}{c}{Baby}           \\
	\cmidrule(lr){2-4} \cmidrule(lr){5-7} \cmidrule(lr){8-10}
	& R@20 & P@20 & NDCG@20 & R@20 & P@20 & NDCG@20 & R@20 & P@20 & NDCG@20 \\ \midrule
	ItemKNN  & 0.0280          & 0.0014          & 0.0131          & 0.0410          & 0.0022          & 0.0212          & 0.0317          & 0.0017          & 0.0152          \\
	MF       & 0.0191          & 0.0010          & 0.0088          & 0.0430          & 0.0023          & 0.0202          & 0.0440          & 0.0024          & 0.0200          \\
	NGCF     & 0.0387          & 0.0020          & 0.0168          & 0.0728          & 0.0038          & 0.0332          & 0.0591          & 0.0032          & 0.0261          \\
	LightGCN & 0.0470          & 0.0024          & 0.0215          & 0.0803          & 0.0042          & 0.0377          & 0.0698          & 0.0037          & 0.0319           \\
	SGL      & 0.0598          & 0.0030          & 0.0268          & \underline{0.0905}          & \underline{0.0047}          & \underline{0.0412}          & 0.0745          & 0.0040          & 0.0328          \\ \midrule
	VBPR     & 0.0481          & 0.0024          & 0.0205          & 0.0582          & 0.0031          & 0.0265          & 0.0486          & 0.0026          & 0.0213          \\
	MMGCN    & 0.0501          & 0.0024          & 0.0221          & 0.0638          & 0.0034          & 0.0279          & 0.0640          & 0.0032          & 0.0284          \\
	GRCN    & \underline{0.0631}         &\underline{0.0032}          & \underline{0.0276}          & 0.0833          & 0.0044          & 0.0377          & \underline{0.0754}          & \underline{0.0040}          & \underline{0.0336}          \\ \midrule
	\rowcolor{gray!20}
	\themodel     & \textbf{0.0782} & \textbf{0.0040} & \textbf{0.0351} & \textbf{0.0988} & \textbf{0.0052} & \textbf{0.0457} & \textbf{0.0892} & \textbf{0.0047} & \textbf{0.0402} \\
	$\Delta$Improvement  & 24.1\%          & 23.1\%          & 27.4\%          & 9.2\%          & 10.6\%          & 10.9\%          & 18.3\%           & 16.9\%           & 19.7\%        \\
	\bottomrule
	\end{tabular}
\end{table*}
	
We conduct experiments on three categories of widely used Amazon datasets introduced by \citet{McAuley:2015ip}: \textsf{(a) Clothing, Shoes and Jewelry}, \textsf{(b) Sports and Outdoors}, and \textsf{(c) Baby}, which we refer to as \textbf{Clothing}, \textbf{Sports} and \textbf{Baby} in brief.
The statistics of these three datasets are summarized in Table \ref{tab:dataset}. The three datasets include both visual and textual modalities. We use the 4,096-dimensional visual features that have been extracted and published. For the textual modality, we extract textual embeddings by concatenating the title, descriptions, categories, and brand of each item and utilize sentence-transformers \cite{Reimers:2019iz} to obtain 1,024-dimensional sentence embeddings.

\subsubsection{Baselines}
To evaluate the effectiveness of our proposed model, we compare it with several state-of-the-art recommendation models. These baselines fall into two groups: CF methods (i.e., ItemKNN, MF, NGCF, LightGCN, SGL) and deep content-aware recommendation models (i.e., VBPR, MMGCN, GRCN).
\begin{itemize}
    \item \textbf{ItemKNN} \cite{deshpande2004item} computes the similarity between the items, and compute the similarity between a basket of items and a candidate recommender item.
	\item \textbf{MF} \cite{Rendle:2009wp} optimizes Matrix Factorization using the Bayesian personalized ranking (BPR) loss, which exploits the user-item direct interactions only as the target value of interaction function.
	\item \textbf{NGCF} \cite{Wang:2019er} explicitly models user-item interactions by a bipartite graph. By leveraging graph convolutional operations, it allows the embeddings of users and items to interact with each other to harvest the collaborative signals as well as high-order connectivity signals.
	\item \textbf{LightGCN} \cite{He:2020gd} argues the unnecessarily complicated design of GCNs (i.e., feature transformation and nonlinear activation) for recommendation systems and proposes a light model which only consists of two essential components: light graph convolution and layer combination.
	\item \textbf{SGL} 	\cite{wu2021self} generates multiple views of a node and maximizing the agreement between different views of the same node.
	\item \textbf{VBPR} \cite{He:2016ww}: Based upon the BPR model, it integrates the visual features and ID embeddings of each item as its representation and feeds them into the Matrix Factorization framework. In our experiments, we concatenate multi-modal features as content information to predict the interactions between users and items.
	\item \textbf{MMGCN} \cite{Wei:2019hn} is one of the state-of-the-art multimodal recommendation methods, which constructs modal-specific graphs and refines modal-specific representations for users and items. It aggregates all model-specific representations to obtain the representations of users or items for prediction.
	\item \textbf{GRCN} \cite{Wei:2020ko} is also one of the state-of-the-arts multimodal recommendation methods. It refines user-item interaction graph by identifying the false-positive feedback and prunes the corresponding noisy edges in the interaction graph.
\end{itemize}

\subsubsection{Evaluation Protocols}
We conduct experiments in both warm-start and cold-start settings.

\textbf{Warm-start settings.}
For each dataset, we select 80\% of historical interactions of each user to constitute the training set, 10\% for validation set, and the remaining 10\% for testing set. For each observed user-item interaction, we treat it as a positive pair, and then conduct the negative sampling strategy to pair them with one negative item that the user does not interact before.

\textbf{Cold-start settings.}
We remove all user-item interaction pairs associated with a randomly selected 20\% item set from the training set. We further divide the half of the items (10\%) into the validation set and half (10\%) into the testing set. In other words, these items are entirely unseen in the training set.

We adopt three widely-used metrics to evaluate the performance of preference ranking: Recall@\(k\), NDCG@\(k\), and Precision@\(k\). By default, we set \(k = 20\) and report the averaged metrics for all users in the testing set.

\subsubsection{Implementation Details}
We implemente our method in PyTorch \cite{Paszke:2019vf} and set the embedding dimension $d$ fixed to 64 for all models to ensure fair comparison. We optimize all models with the Adam \cite{Kingma:2015us} optimizer, where the batch size is fixed at 1024. We use the Xavier initializer \cite{Glorot:2010uc} to initialize the model parameters. The optimal hyper-parameters are determined via grid search on the validation set: the learning rate is set to $0.0005$, the coefficient of $\ell_2$ normalization is set to $10^{-4}$. The $k$ of $k$NN sparsification is set to $10$, the $\lambda$ of skip connection is set to $0.7$, the temperature parameter $\tau$ is set to $0.5$, the coefficient $\beta$ used to control the effect of contrastive auxiliary task is set to $0.03$. Besides, we stop training if Recall@20 on the validation set does not increase for 10 successive epochs to avoid overfitting.

\begin{table*}
	\centering
    \caption{Performance of our proposed \themodel on top of different downstream collaborative filtering (CF) methods. $\Delta$Improvement indicates the relative improvements of \themodel compared to the variant \textbf{CF+feats} in percentage.}
    \label{tab:ablation}
    \begin{tabular}{cccccccccc}
    \toprule
    \multirow{2.5}{*}{Model} & \multicolumn{3}{c}{Clothing}       & \multicolumn{3}{c}{Sports}         & \multicolumn{3}{c}{Baby}           \\ \cmidrule(lr){2-4} \cmidrule(lr){5-7} \cmidrule(lr){8-10}
    & R@20 & P@20 & NDCG@20 & R@20 & P@20 & NDCG@20 & R@20 & P@20 & NDCG@20 \\
    \midrule
    MF             & 0.0191    & 0.0010       & 0.0088  & 0.0430               & 0.0023               & 0.0202               & 0.0440               & 0.0024               & 0.0200               \\
    MF+feats       & 0.0456    & 0.0023       & 0.0197  & 0.0674               & 0.0036               & 0.0304               & 0.0701               & 0.0037               & 0.0306               \\
    \themodel/feats       & 0.0627    & 0.0032       & 0.0276  & 0.0823               & 0.0043               & 0.0372               & 0.0766               & 0.0040               & 0.0334               \\
    \themodel w/o. contrast       & 0.0664    & 0.0034       & 0.0301  & 0.0853               & 0.0045               & 0.0397                & 0.0701               & 0.0036               & 0.0309               \\
    \rowcolor{gray!20}
    \themodel        & 0.0715    & 0.0036       & 0.0319  & 0.0875               & 0.0046               & 0.0402               & 0.0758               & 0.0040               & 0.0337               \\
    $\Delta$Improvement       & 56.8\%    & 56.5\%       & 61.9\%   & 29.8\%   & 27.8\%   & 32.3\%   & 8.1\%   & 8.1\%   & 10.1\%  \\
    \midrule
    NGCF           & 0.0387    & 0.0020       & 0.0168  & 0.0728               & 0.0038               & 0.0332               & 0.0591               & 0.0032               & 0.0261               \\
    NGCF+feats     & 0.0436    & 0.0022       & 0.0190  & 0.0748               & 0.0040               & 0.0344               & 0.0660               & 0.0035               & 0.0295               \\
    \themodel/feats       & 0.0539    & 0.0027       & 0.0235  & 0.0796               & 0.0042               & 0.0359               & 0.0751               & 0.0040               & 0.0324               \\
    \themodel w/o. contrast       & 0.0486    & 0.0025       & 0.0213  & 0.0804               & 0.0043               & 0.0360               & 0.0700               & 0.0037               & 0.0307               \\
    \rowcolor{gray!20}
    \themodel      & 0.0637    & 0.0032       & 0.0279  & 0.0874               & 0.0046               & 0.0396               & 0.0800               & 0.0042               & 0.0351               \\
    $\Delta$Improvement       & 46.1\%    & 45.5\%       & 46.8\%   & 16.8\%   & 15.0\%   & 15.1\%   & 21.2\%   & 20.0\%   & 19.0\%  \\
    \midrule
    LightGCN       & 0.0470    & 0.0024       & 0.0215  & 0.0803               & 0.0042               & 0.0377               & 0.0698               & 0.0037               & 0.0319               \\
    LightGCN+feats & 0.0477    & 0.0024       & 0.0208  & 0.0754               & 0.0040               & 0.0350               & 0.0793               & 0.0042               & 0.0344               \\
    \themodel/feats       & 0.0659    & 0.0033       & 0.0289  & 0.0889               & 0.0047               & 0.0406               & 0.0871               & 0.0046               & 0.0388               \\    
    \themodel w/o. contrast       & 0.0703    & 0.0036       & 0.0321  & 0.0925               & 0.0049               & 0.0428               & 0.0830               & 0.0044               & 0.0359               \\
    \rowcolor{gray!20}
    \themodel  & 0.0782    & 0.0040       & 0.0351  & 0.0988               & 0.0052               & 0.0457               & 0.0892               & 0.0047               & 0.0402               \\
    $\Delta$Improvement       & 63.9\%    & 66.7\%       & 68.8\%   & 31.0\%   & 30.0\%   & 30.6\%   & 12.5\%   & 11.9\%   & 16.9\%  \\
    \bottomrule
    \end{tabular}
\end{table*}

\subsection{Performance Comparison (RQ1)}
\label{sec:rq1}
We start by comparing the performance of all methods, and then explore how the our method alleviate the cold-start problem. In this subsection, we combine \themodel with LightGCN as a downstream CF method, and will also conduct experiments with different CF methods in Section \ref{sec:rq2}.

\subsubsection{Overall Performances}
\label{sec:overall-perf}
Table \ref{tab:experiments} reports the performance comparison results, from which we can observe:

\begin{itemize}
	\item Our method \themodel significantly outperforms both CF methods and content-aware methods, verifying the effectiveness of our methods. Specifically, \themodel improves over the strongest baselines in terms of Recall@20 by 24.1\%, 18.6\%, and 18.3\% in Clothing, Sports, and Baby, respectively. This indicates that our proposed method is well-designed for multimedia recommendation by discovering underlying item-item relationships and conducting fine-grained multimodal fusion through the contrastive auxiliary task.
	\item Compared with CF methods, content-aware methods yield better performance overall, which indicates that multimodal features provide rich content information about items, and can boost recommendation accuracy. Even without utilizing item content information, the self-supervised method SGL achieves competitive performance on the three datasets and even outperforms the powerful content-aware method GRCN, since the auxiliary self-supervised task can augment the node representation learning
	\item Additionally, existing content-aware recommendation models are highly dependent on the representativeness of multimodal features and thus obtain fluctuating performances over different datasets. For Clothing dataset where visual features are very important in revealing item attributes \cite{Liu:2017ij,He:2016ww}, VBPR, MMGCN, and GRCN outperform all CF methods. For the other two datasets where multimodal features may not directly reveal item attributes, content-aware methods obtain relatively small improvements. The performances of VBPR and MMGCN are even inferior to the CF method LightGCN. Different from existing content-aware methods, we discover the latent item relationships underlying multimodal features instead of directly using them as side information. The latent item relationships are less dependent on the representativeness of multimodal features, and thus we are able to obtain robust performance.
\end{itemize}

\subsubsection{Performances in Cold-start Settings}
\label{sec:cold-start}
\begin{figure}
	\centering
	\includegraphics[width=\linewidth]{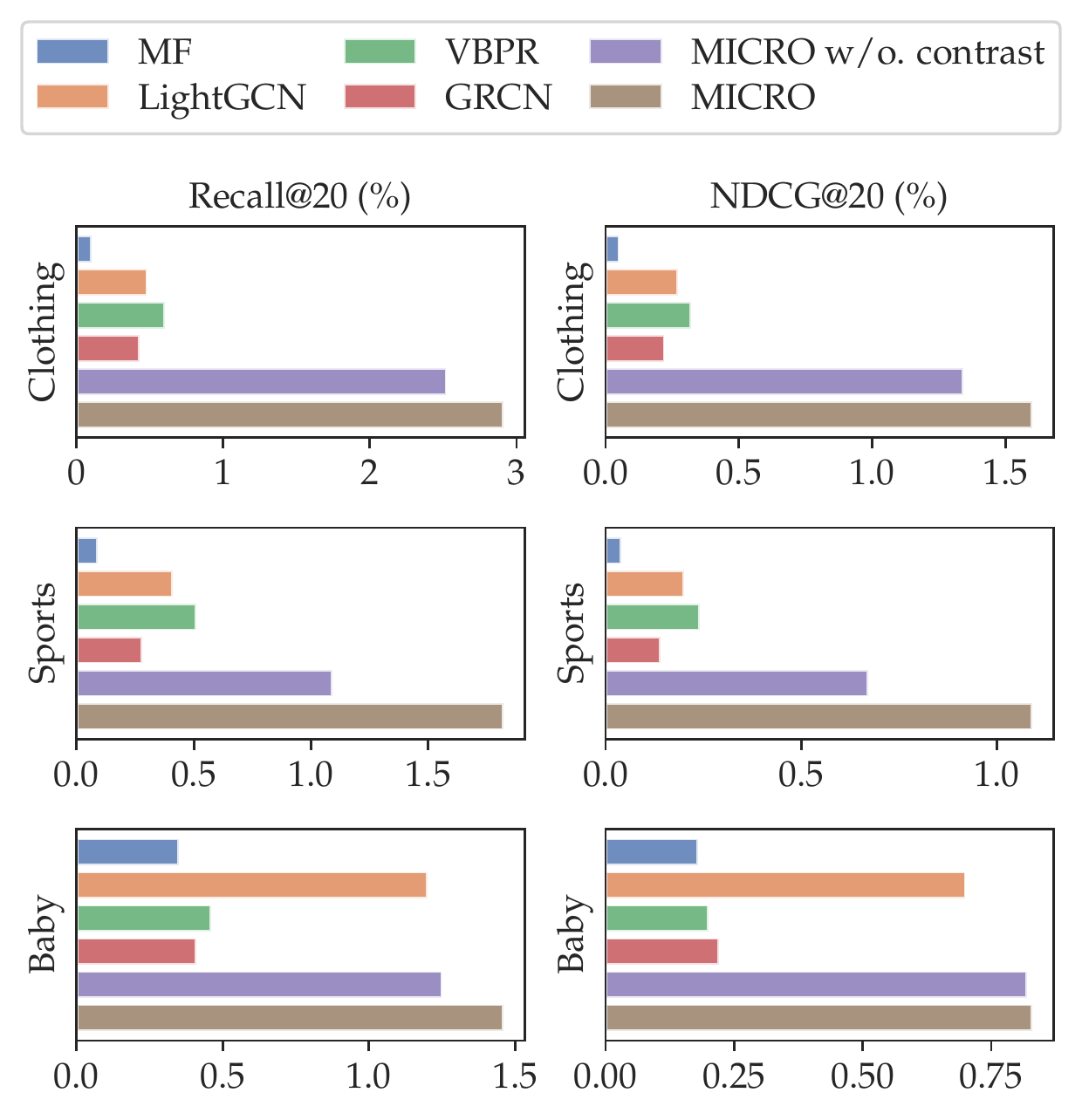}
	\caption{Performance comparison of our method with different baselines in cold-start settings.}
	\label{fig:cold-start}
\end{figure}

The cold-start problem remains a prominent challenge in recommendation systems \cite{Schein:2002gw}. Multimodal features of items provide rich content information, which can be exploited to alleviate the cold-start problem. We conduct cold-start experiments and compare with representative baselines. \textbf{\themodel w/o. contrast} is the simplified variant of \themodel, which discards the contrastive auxiliary task in Section \ref{sec:contrast} and only utilizes the BPR loss in Eq. (\ref{eq:bpr}). Figure \ref{fig:cold-start} reports the results of performance comparison, from which we can observe:
\begin{itemize}
	\item Both \themodel w/o. contrast and \themodel can alleviate the cold-start problem and outperform all baselines on three datasets. They learn item graphs from multimodal features, along which cold-start items will get similar feedbacks from relevant neighbors through neighborhood aggregation of graph convolutions.
	\item Additionally, \themodel outperforms \themodel w/o. contrast on three datasets. In \themodel, the multimodal contrastive framework serves as a self-supervised auxiliary task. The self-supervision signals are constructed by maximizing the agreement between item representations under individual modalities and the multimodal fused representations to learn better item representations which encode item relationships from multiple modalities. In this way, the cold-start problem would be further alleviated.
	\item CF methods MF and LightGCN obtain poor performances under cold-start settings in general, primarily because they only leverage users’ feedbacks to predict the interactions between users and items. Although these methods may work well for items with sufficient feedbacks, they cannot help in cold-start settings, since no user-item interaction is available to update the representations of cold-start items.
	\item Content-aware model VBPR outperforms CF methods in general, which indicates that the content information provided by multimodal features benefits recommendation for cold-start items. In particular, content information can help bridge the gap between the existing items to cold-start items. However, some graph-based content-aware methods such as GRCN, although perform well in warm-start settings, obtain poor performance in cold-start settings. GRCN utilizes multimodal features on user-item interaction bipartite graphs, which is also heavily dependent on user-item interactions. For cold-start items, they never interact with users and become isolated nodes in the user-item graphs, leading to inferior performance.
\end{itemize}

\subsection{Ablation Studies (RQ2)}
\label{sec:rq2}
In this subsection, we combine \themodel with three common-used CF methods, i.e., MF, NGCF, and LightGCN to validate the effectiveness and flexibility of our proposed method. For each CF method, we have three other variants: the first one is \textbf{CF+feats}, which does not consider latent item-item relationships and directly uses transformed multimodal features to replace the item representations learned from item graphs in Eq. (\ref{eq:cf}). The second one is named as \textbf{\themodel/feats}, which uses multimodal features as the input initial item embeddings of graph convolutions instead of ID embeddings. The third is \textbf{\themodel w/o. contrast}, which discards the contrastive auxiliary task in Section \ref{sec:contrast} and only utilizes the BPR loss in Eq. (\ref{eq:bpr}). Table \ref{tab:ablation} summarizes the performance and the relative improvements gained by \textbf{\themodel} over \textbf{CF+feats}, from which we have the following observations:
\begin{itemize}
	\item {\themodel} significantly and consistently outperforms all original CF methods and {CF+feats} variants on three datasets, obtaining up to 68.8\% improvements over the {CF+feats} variants, verifying the flexibility of our plug-in paradigm. 
	\item Even without the contrastive auxiliary task, {\themodel w/o. contrast} obtains significant improvements over {CF+feats}, indicating the effectiveness of discovering latent item-item relationships from multimodal features. Furthermore, the improvements between {\themodel} and {\themodel w/o. contrast} show the importance of fine-grained multimodal fusion, through which we can capture item relationships shared between modalities adaptively.
	\item Based on the learned item graph structures, {\themodel/feats} employs graph convolutions on multimodal features. Our proposed method {\themodel} utilizes the same learned structures but employ graph convolutions on item ID embeddings, which aims to directly model \emph{item affinities}. The improvements between them validate the effectiveness of explicitly modeling item affinities where multimodal features are only used to bridge semantic relationships between items.
\end{itemize}

\subsection{Sensitivity Analysis (RQ3)}
\label{section:rq3}

\begin{figure}
	\centering
	\includegraphics[width=\linewidth]{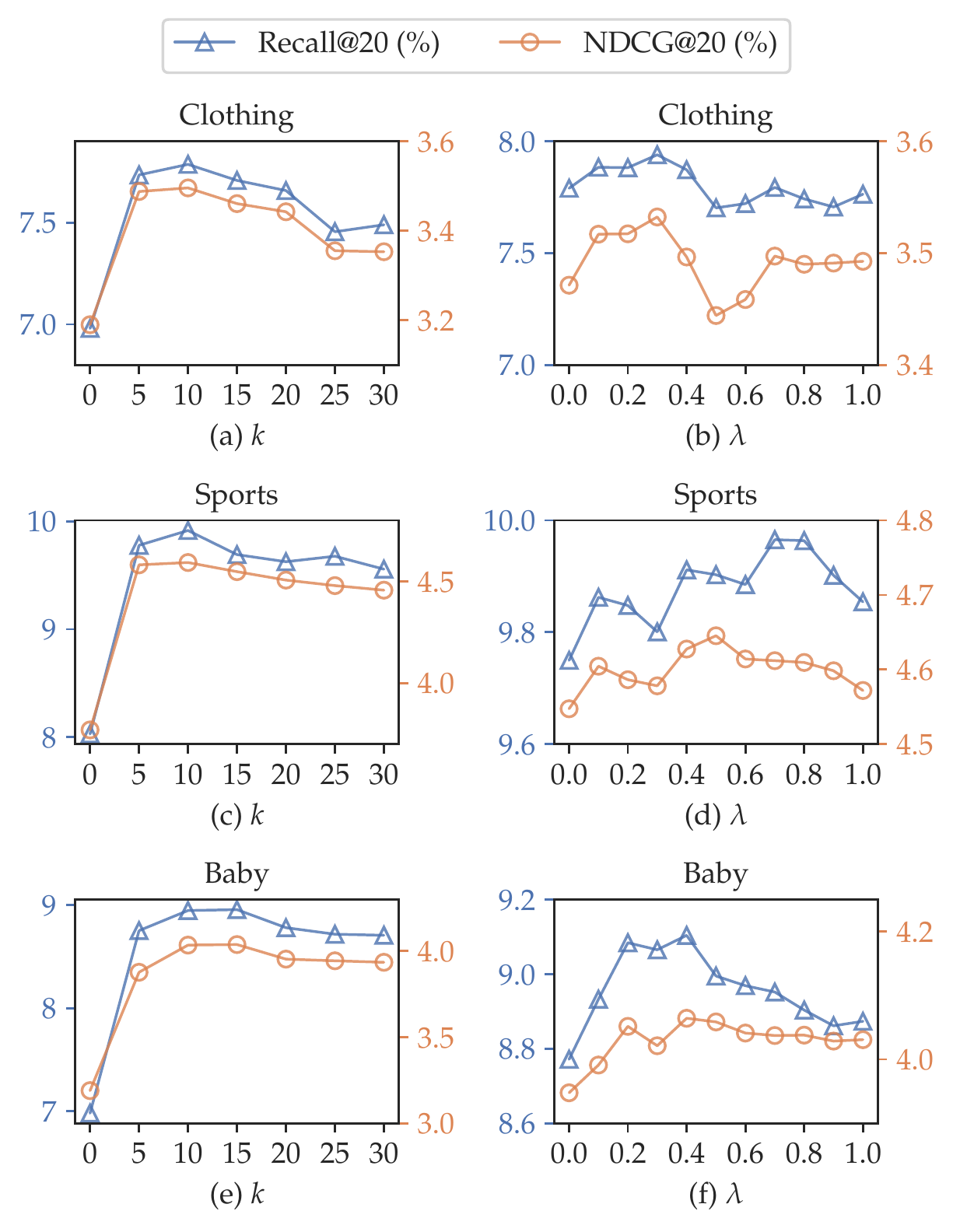}
	\caption{Performance comparison of various hyperparameters \(k\) and \(\lambda\).}
	\label{fig:hyperparams}
\end{figure}

Since the graph structure learning layer and the contrastive auxiliary task play pivotal roles in our method, in this subsection, we conduct sensitivity analysis with different hyper-parameters on graph structure learning layers and the contrastive auxiliary task. Firstly, we investigate performance of \themodel-LightGCN with respect to different $k$ value of the $k$-NN sparsification operation since $k$ is important which determines the number of neighbors of each item, and controls the amount of information propagated from neighbors. Secondly, we discuss how the skip connection coefficient $\lambda$ affects the performance which controls the amount of information from the initial graph structures. Finally, we explore how the contrastive auxiliary task magnitude $\beta$ affects the performance.

\subsubsection{Impact of Varied $k$ Values}
Figures \ref{fig:hyperparams}(a)(c)(e) present the results of performance comparison. $k=0$ means no item relationships are included and the model is degenerated to LightGCN. We have the following observations:
\begin{itemize}
	\item Our method gains significant improvement between $k=0$ and $k=5$, which validates the rationality of item relationships mined from multimodal features. Even if only a small part of the neighbors are included, we can obtain better item representations by aggregating meaningful and important information from the neighbors, which boost the recommendation performance.
	\item Furthermore, the performance first improves as $k$ increases, which verifies the effectiveness of information aggregation along item-item graphs since more neighbors bring more meaningful information that helps to make more accurate recommendations.
	\item The trend, however, declines when $k$ continues to increase, since there may exist many unimportant neighbors that will inevitably introduce noise to information propagation. This demonstrates the necessity of conducting $k$NN sparsification on the learned dense graph.
\end{itemize}

\subsubsection{Impact of Varied Coefficients $\lambda$}
Figures \ref{fig:hyperparams}(b)(d)(f) present the performance comparison. $\lambda=0$ means only consider the graph structure learned by the transformed multimodal features, and $\lambda=1$ means we only consider the initial structure generated by the raw multimodal features. We have the following observations:
\begin{itemize}
	\item When we set $\lambda = 0$, the model obtains poor performance. It only learns graph structure from the transformed features, completely updating the adjacency matrix every time, ignoring the rich and useful information of raw features and resulting in fluctuated training process.
	\item The performance first grows as $\lambda$ becomes larger, validating the importance of initial structures constructed by raw multimodal features. However, it begins to deteriorate when $\lambda$ continues to increase, since raw features are often noisy due to the inevitably error-prone data measurement or collection process. Learning the graph structures dynamically can reduce noise. 
	\item Overall, there are no apparent sharp rises and falls, indicating that our method is not that sensitive to the selection of $\lambda$. Notably, all models surpass the baselines (c.f. Table \ref{tab:experiments}), proving the effectiveness of item graphs.
\end{itemize}

\begin{figure}
	\centering
	\includegraphics[width=\linewidth]{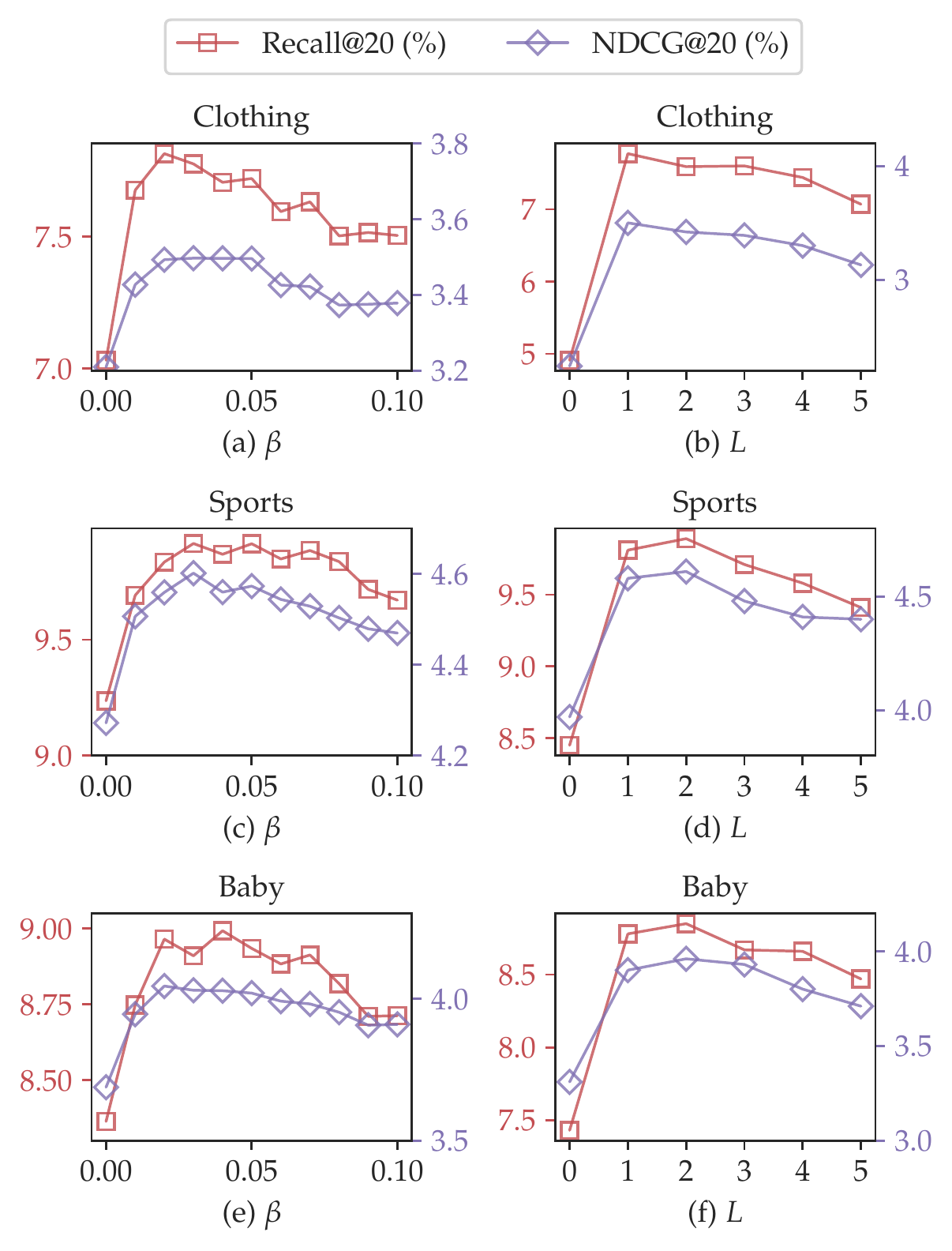}
	\caption{Performance comparison of various hyperparameters \(\beta\) and \(L\).}
	\label{fig:hyperparams-2}
\end{figure}

\subsubsection{Impact of Varied Coefficients $\beta$}
We investigate how the contrastive auxiliary task magnitude $\beta$ affects the performance. Figures \ref{fig:hyperparams-2}(a)(c)(e) report the performance comparison. $\beta=0$ denotes \themodel w/o. contrast, which discards the contrastive auxiliary task. We can observe that 
\begin{itemize}
    \item With the increase of $\beta$, the performances on all datasets first rise and is always better than $\beta=0$. The primary recommendation task achieves decent gains when jointly optimized with the self-supervised auxiliary task even with a small $\beta$.
    \item  However, it begins to decline when $\beta$ continues to increase. A small $\beta$ can promote the primary task, while a larger one would mislead it. The benefits brought by the self-supervised task could be easily neutralized and the recommendation task is sensitive to the magnitude of the self-supervised task, which is also observed in other recommendation works with contrastive learning \cite{yu2021self, xia2021self}.
\end{itemize}

\subsubsection{Impact of Varied Layer Number $L$}
In order to investigate the effect of multiple graph convolution layers and high-order information, we search the number of layers $L$ in the range of $\{0, 1, 2, 3, 4, 5\}$. Figures \ref{fig:hyperparams-2}(b)(d)(f) report the performance comparison. We can observe that 
\begin{itemize}
    \item When $L$ increases from 0 to 1, the performance increases significantly on all datasets, indicating that the item-item relationships can effectively boost recommendation.
    \item The best performed hop varies from different datasets. Specifically, \themodel achieves the best performance with $L=1$ in Clothing, $L=2$ in Sports and Baby. Applying a too deep architecture might introduce noisy, unimportant item relationships to the representation learning.
    \item When varying the number of layers, \themodel  consistently and significantly outperforms baselines on all datasets. It again verifies the effectiveness of item-item relationships.
\end{itemize}

\subsection{Investigation of the Contribution of Each Modality (RQ4)}
\begin{table}
\centering
\caption{Performances comparison over different modalities.}
\label{tab:modal}
\begin{tabular}{@{}ccccc@{}}
\toprule
Dataset & Model                    & R@20 & P@20 & NDCG@20  \\ \midrule
\multirow{3}{*}{Clothing} 
& Visual               &  0.0613         &   0.0031  &   0.0282    \\
& Textual                &  0.0699         &   0.0035  &   0.0321    \\ 
& Both               &  0.0782         &   0.0040  &   0.0351    \\   \midrule
\multirow{3}{*}{Sports} 
& Visual               &  0.0879         &   0.0046  &   0.0408    \\
& Textual                &  0.0928         &   0.0048  &   0.0430    \\ 
& Both               &  0.0988         &   0.0052  &   0.0457    \\   \midrule
\multirow{3}{*}{Baby} 
& Visual               &  0.0772         &   0.0040  &   0.0348    \\
& Textual                &  0.0825         &   0.0043  &   0.0373    \\ 
& Both               &  0.0892         &   0.0047  &   0.0402    \\   \bottomrule
\end{tabular}
\end{table}
In this subsection, we aim to explore the contribution of each modality. Table \ref{tab:modal} reports the performance comparison over different modalities. We observe that the performances of utilizing multiple modalities are better than that of ones within the single modality, demonstrating that incorporating the information from multiple modalities facilitates comprehensive understanding of items. Additionally, textual modality contributes more than visual modality in general. It is reasonable since textual modality provide more fine-grained information which directly reveals the titles, categories and descriptions of items while visual modality only provides coarse-grained visual appearances.

\section{Related Work}
\subsection{Multimedia Recommendation}
Collaborative filtering (CF) has achieved great success in recommendation systems, which leverage users' feedbacks (such as clicks and purchases) to predict the preference of users and make recommendations.
However, CF-based methods suffer from sparse data with limited user-item interactions and rarely accessed items. To address the problem of data sparsity, it is important to exploit other information besides user-item interactions.
Multimedia recommendation systems consider massive multimedia content information of items, which have been successfully applied to many applications, such as e-commerce, instant video platforms and social media platforms \cite{veit2015learning,McAuley:2015ip,He:2016dm,Cui:2021ks}.

For example, VBPR \cite{He:2016ww} extends matrix factorization by incorporating visual features extracted from product images to improve the performance.
DVBPR \cite{Kang:2017ds} attempts to jointly train the image representation as well as the parameters in a recommender model.
Sherlock \cite{he2016sherlock} incorporates categorical information for recommendation based on visual features.
DeepStyle \cite{Liu:2017ij} disentangles category information from visual representations for learning style features of items and sensing preferences of users.
ACF \cite{Chen:2017jj} introduces an item-level and component-level attention model for inferring the underlying users' preferences encoded in the implicit user feedbacks. 
VECF \cite{Chen:2019ga} models users' various attentions on different image regions and reviews.
MV-RNN \cite{cui2018mv} uses multimodal features for sequential recommendation in a recurrent framework.
Recently, Graph Neural Networks (GNNs) have been introduced into recommendation systems \cite{Wu:2019ke,Wang:2019er,zhang2020personalized} and especially multimodal recommendation systems \cite{Wei:2019hn,Wei:2020ko,li2020hierarchical}. MMGCN \cite{Wei:2019hn} constructs modal-specific graph and conducts graph convolutional operations, to capture the modal-specific user preference and distills the item representations simultaneously. In this way, the learned user representation can reflect the users’ specific interests on items. Following MMGCN, GRCN \cite{Wei:2020ko} focuses on adaptively refining the structure of interaction graph to discover and prune potential false-positive edges. There are several prior studies \cite{park2017also, Lee2017LargeScaleCV, wei2021hierarchical} that propose to explore collaborative item relationships through high-order item-user-item co-occurrences. For example, HUIGN \cite{wei2021hierarchical} constructs a co-interacted item graph which exhibits users’ intents at different levels. It aims to learn multi-level user intents from the co-interacted patterns of items and further enhance the recommendation performance.

The above methods directly utilize multimodal features as side information of each item and disregard fine-grained multimodal fusion. In our model, we step further by discovering semantic item-item relationships from multimodal features, and conduct  fine-grained  multimodal fusion to inject complementary item-item relationships from multiple modalities into the item representations.

\subsection{Deep Graph Structure Learning}
GNNs have shown great power on analyzing graph-structured data and have been widely employed for graph analytical tasks across a variety of domains, including node classification \cite{Kipf:2017tc,Zhu:2020vf}, link prediction \cite{Chen:2018vh}, information retrieval\cite{Zhang:2021vf, Yu:2021ka}, etc.
However, most GNN methods are highly sensitive to the quality of graph structures and usually require a perfect graph structure that are hard to construct in real-world applications \cite{Franceschi:2019uz}. Since GNNs recursively aggregate information from neighborhoods of one node to compute its node embedding, such an iterative mechanism has cascading effects --- small noise in a graph will be propagated to neighboring nodes, affecting the embeddings of many others. Additionally, there also exist many real-world applications where initial graph structures are not available. Recently, considerable literature has arisen around the central theme of Graph Structure Learning (GSL), which targets at jointly learning an optimized graph structure and corresponding representations. There are three categories of GSL methods: metric learning \cite{Chen:2020wu,Wang:2020bs,Li:2018wu}, probabilistic modeling \cite{Franceschi:2019uz,Zheng:2020tp,Luo:2021gg}, and direct optimization approaches \cite{Yang:2019fh,Jin:2020br,Gao:2020em}. 

For example, IDGL \cite{Chen:2020wu} casts the graph learning problem into a similarity metric learning problem and leverage adaptive graph regularization for controlling the quality of the learned graph; DGM \cite{Kazi:2020vj} predicts a probabilistic graph, allowing a discrete graph to be sampled accordingly in order to be used in any graph convolutional operator. NeuralSparse \cite{Zheng:2020tp} considers the graph sparsification task by removing task-irrelevant edges. It utilizes a deep neural network to learn $k$-neighbor subgraphs by selecting at most $k$ neighbors for each node in the graph.  We kindly refer to \cite{Zhu:2021ue} for a recent overview of approaches for graph structure learning.

In personalized recommendation, although user-item interactions can be formulated as a bipartite graph naturally, item-item relations remain rarely explored. To model item relationships explicitly, we employ metric learning approaches to represent edge weights as a distance measure between two end nodes, which fits for multimedia recommendation since rich content information can be included to measure the semantic relationship between two items.

\subsection{Contrastive Learning}
Self-supervised learning is an emerging technique to learn representations by self-defined supervision signals generated from raw data without relying on annotated labels. Contrastive learning (CL) has become a dominant branch of self-supervised learning, which targets at obtaining robust and discriminative representations by grouping positive samples closer and negative samples far from each other. For visual data, negative samples can be generated using a multiple-stage augmentation pipeline \cite{Chen:2020wj,Bachman:2019wp,Falcon:2020uv}, consisting of color jitter, random flip, cropping, resizing, rotation, color distortion, etc. The latest advances extend self-supervised learning to graph representation learning. \citet{velickovic2019deep} introduce an objective function measuring the Mutual Information (MI) between global graph embeddings and local node embeddings. GraphCL \cite{you2020graph} and GRACE \cite{Zhu:2020vf} propose a node-level contrastive objective to simplify previous work. Furthermore, \citet{Zhu:2021gx} propose a contrastive method with adaptive augmentation that incorporates various priors for topological and semantic aspects of the graph. Generally, most CL work differs from each other in terms of the generation of negative samples and contrastive objectives. 

There also exist several works combining self-supervised learning with collaborative filtering \cite{wu2021self, yao2021self}, session-based recommendation \cite{zhou2020s3}, social recommendation \cite{long2021social, yu2021self} and multimedia recommendation \cite{liu2020pre, wei2021contrastive}. \citet{wu2021self} introduce self-supervised auxiliary task into collaborative filtering and improve both accuracy and robustness of GNNs for recommendation. \citet{yao2021self} utilize self-supervised learning to learn better latent relationship of item features for large-scale item recommendations. \citet{zhou2020s3} utilize contrastive learning to learn the correlations among attribute, item, subsequence, and sequence. \citet{wei2021contrastive} aim to maximize the mutual information between item content and collaborative signals to alleviate the cold-start problem.

In this work, since multiple modality-aware graphs are involved, the individual modality-aware item representations and multimodal fused representations are natural positive pairs. We utilize contrastive learning to \emph{maximize the agreement between item representations under an individual modality and the multimodal fused representations}. In this way, the fused multimodal representations can adaptively capture item-item relationships shared between multiple modalities in a self-supervised manner. 

\section{Conclusion}
In this paper, we have proposed the latent structure mining method (\themodel) for multimodal recommendation, which leverages graph structure learning to discover latent item relationships underlying multimodal features and devises a novel contrastive framework to fuse multimodal item relationships. In particular, we first develop a modality-aware structure learning layer and graph convolutions to inject modality-aware item relationships into item representations. Furthermore, we propose a novel multimodal contrastive framework to adaptively capture item-item relationships shared between multiple modalities in a self-supervised manner. Finally, the resulting enhanced item representations are infused with item relationships in multiple modalities, which will be added into the output item embeddings of CF models to make recommendations. Empirical results on three public datasets have demonstrated the effectiveness of our proposed model.

\section*{Acknowledgments}
This work was supported by National Key Research and Development Program (2018YFB1402600), National Natural Science Foundation of China (61772528), Beijing National Natural Science Foundation (4182066), Shandong Provincial Key Research and Development Program (2019JZZY010119) and CCF-AFSG Research Fund (20210001).

\bibliographystyle{IEEEtranN}
{
\small
\bibliography{tkde}
}

\begin{IEEEbiography}[{\includegraphics[width=1in,height=1.25in,clip,keepaspectratio]{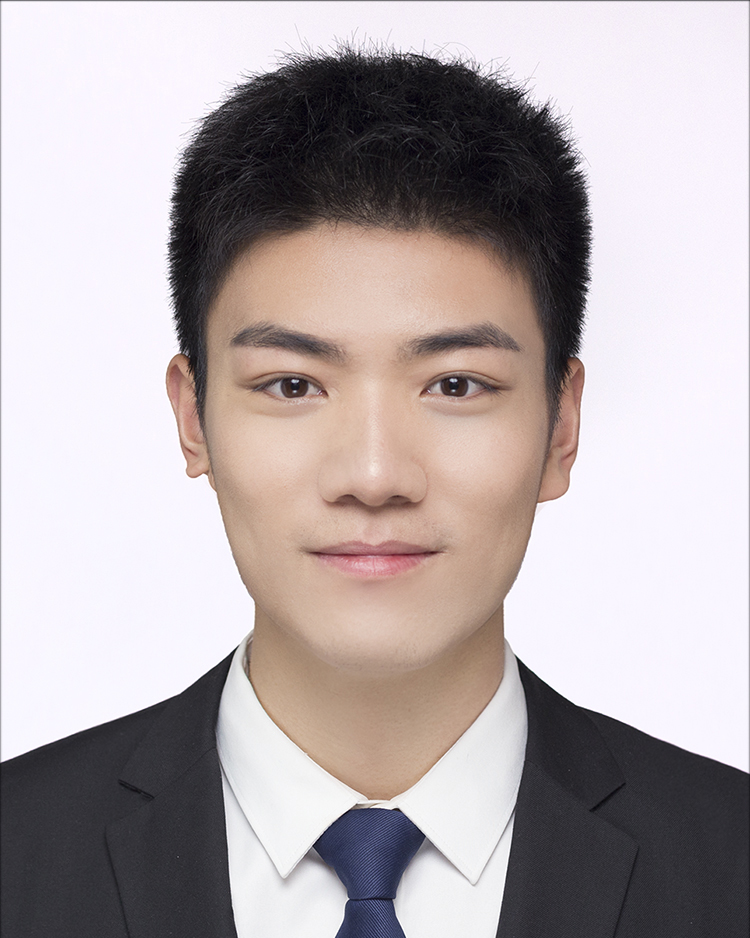}}]
{Jinghao Zhang} is currently pursuing his Ph.D. degree of Computer Science at the Center for Research on Intelligent Perception and Computing (CRIPAC) at National Laboratory of Pattern Recognition (NLPR), Institute of Automation, Chinese Academy of Sciences (CASIA). His current research interests mainly include graph representation learning and recommender systems.
\end{IEEEbiography}

\begin{IEEEbiography}[{\includegraphics[width=1in,height=1.25in,clip,keepaspectratio]{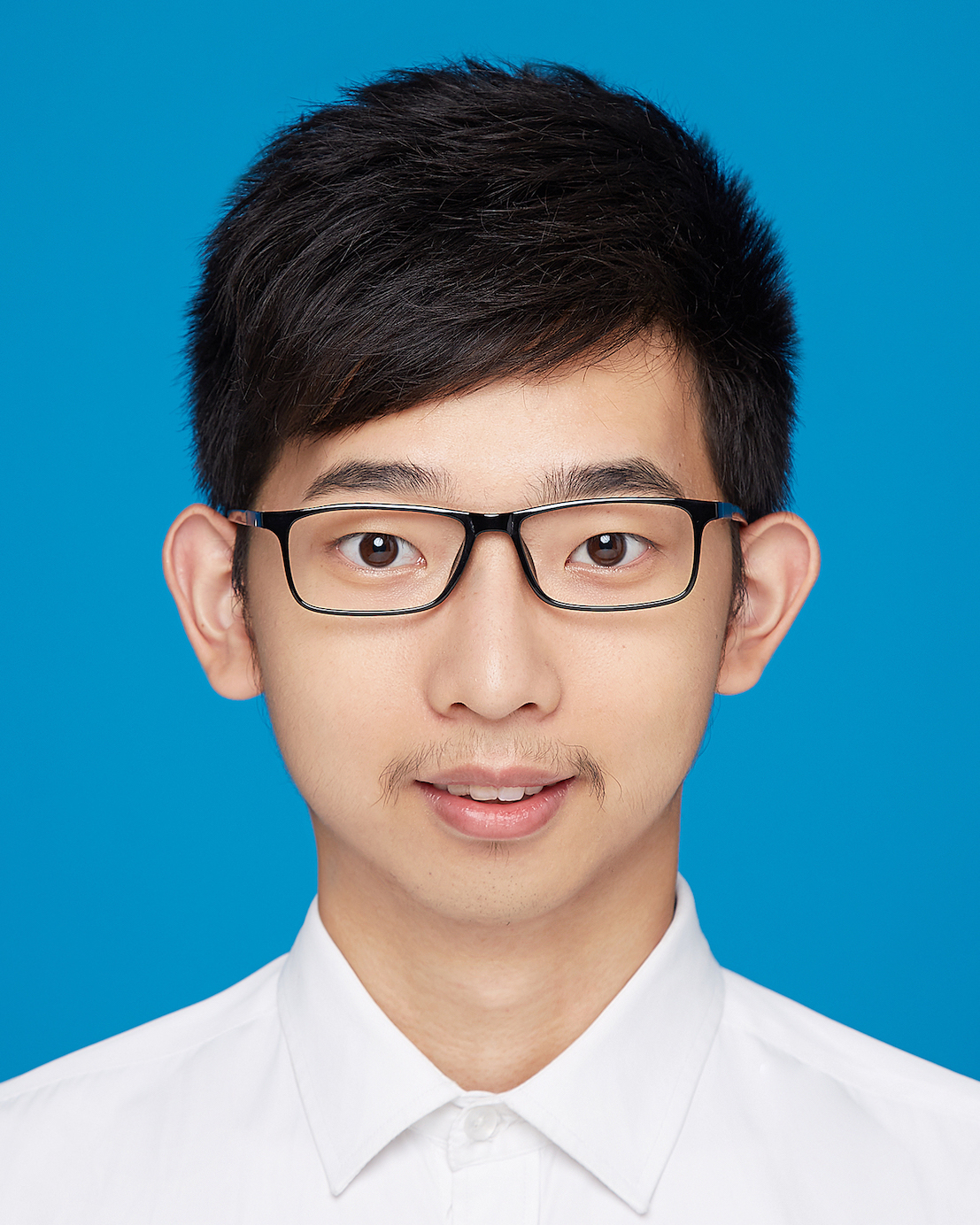}}]
{Yanqiao Zhu} is currently pursuing his master's degree of Computer Science at the Center for Research on Intelligent Perception and Computing (CRIPAC) at National Laboratory of Pattern Recognition (NLPR), Institute of Automation, Chinese Academy of Sciences (CASIA). His current research interests mainly lie in the fields of machine learning with an emphasis on graph representation learning and its application to recommender systems.
\end{IEEEbiography}

\begin{IEEEbiography}[{\includegraphics[width=1in,height=1.25in,clip,keepaspectratio]{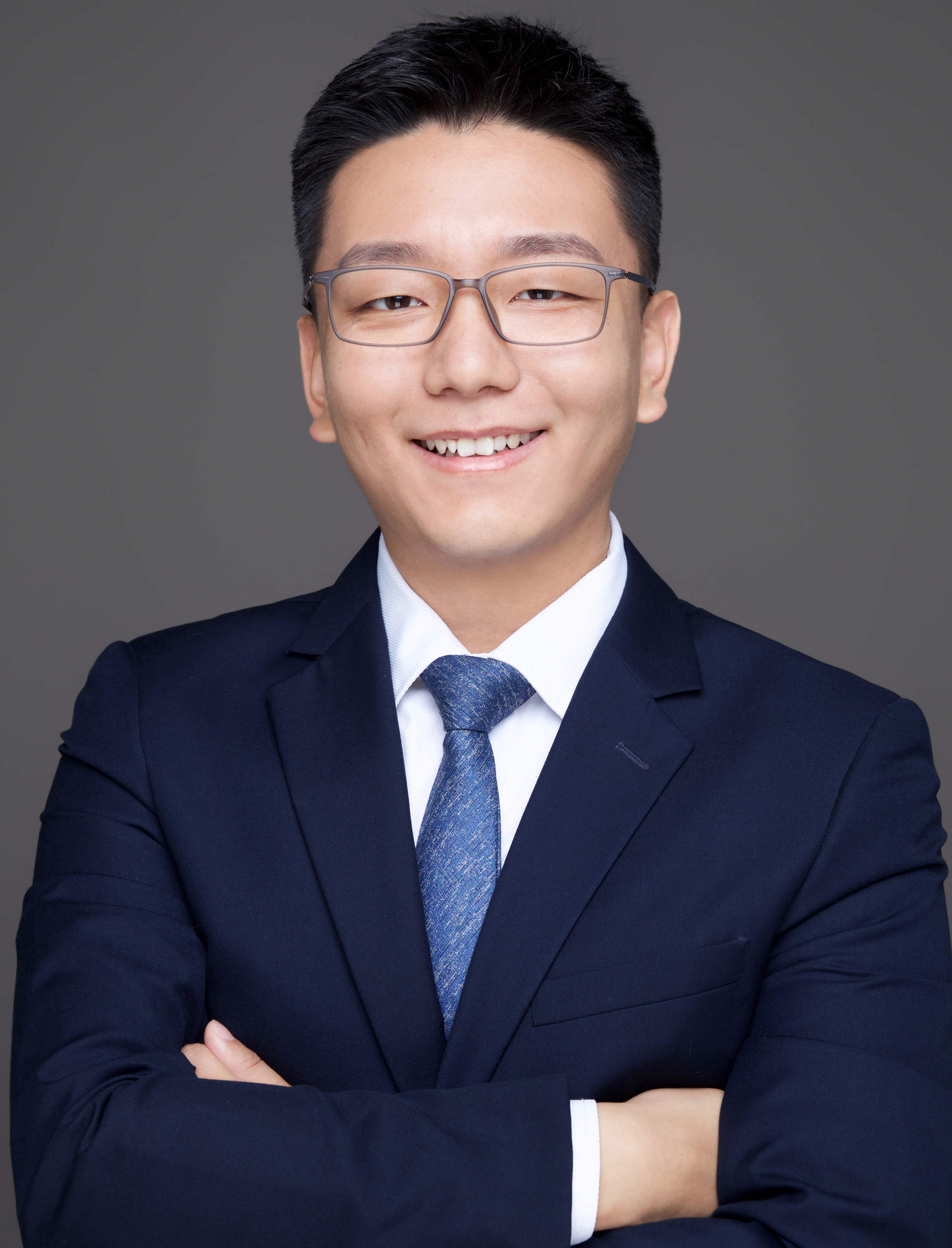}}]
{Qiang Liu} is an Assistant Professor with the Center for Research on Intelligent Perception and Computing (CRIPAC), Institute of Automation, Chinese Academy of Sciences (CASIA). He received his PhD degree in pattern recognition from CASIA. Currently, his research interests include data mining, recommender systems, text mining, knowledge graph, graph representation learning, and causal inference. He has published more than 30 papers in top-tier journals and conferences, such as IEEE TKDE, AAAI, IJCAI, NeurIPS, WWW, SIGIR, CIKM and ICDM.
\end{IEEEbiography}

\begin{IEEEbiography}[{\includegraphics[width=1in,height=1.25in,clip,keepaspectratio]{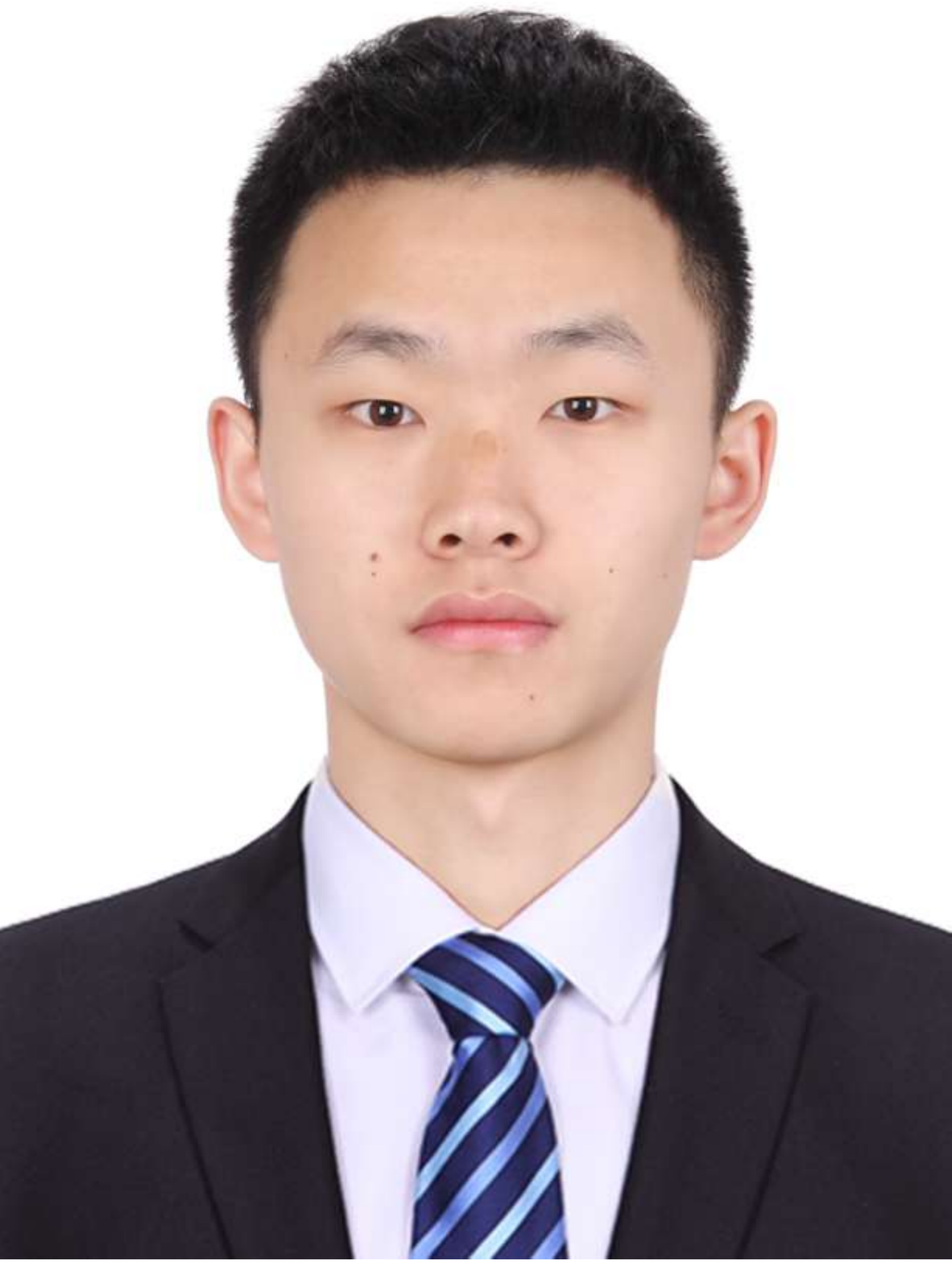}}]
{Mengqi Zhang} is currently pursuing the Ph.D. degree in computer application technology with the University of Chinese Academy of Sciences, Beijing, China. His research interests include data mining, graph representation learning, and recommender systems.
\end{IEEEbiography}

\begin{IEEEbiography}[{\includegraphics[width=1in,height=1.25in,clip,keepaspectratio]{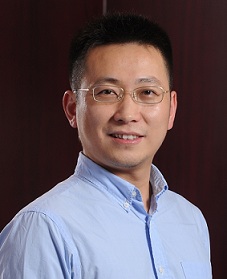}}]
{Shu Wu} received his B.S. degree from Hunan University, China, in 2004, M.S. degree from Xiamen University, China, in 2007, and Ph.D. degree from Department of Computer Science, University of Sherbrooke, Quebec, Canada, all in computer science. He is an Associate Professor with the Center for Research on Intelligent Perception and Computing (CRIPAC) at National Laboratory of Pattern Recognition (NLPR), Institute of Automation, Chinese Academy of Sciences (CASIA). He has published more than 50 papers in the areas of data mining and information retrieval in international journals and conferences, such as IEEE TKDE, IEEE THMS, AAAI, ICDM, SIGIR, and CIKM. His research interests include data mining, information retrieval, and recommendation systems.
\end{IEEEbiography}

\begin{IEEEbiography}[{\includegraphics[width=1in,height=1.25in,clip,keepaspectratio]{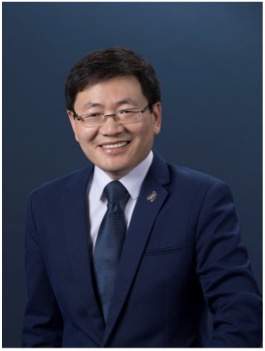}}]
{Liang Wang} received both the BEng and MEng degrees from Anhui University in 1997 and 2000, respectively, and the PhD degree from the Institute of Automation, Chinese Academy of Sciences (CASIA) in 2004. From 2004 to 2010, he was a research assistant at Imperial College London, United Kingdom, and Monash University, Australia, a research fellow at the University of Melbourne, Australia, and a lecturer at the University of Bath, United Kingdom, respectively. Currently, he is a full professor of the Hundred Talents Program at the National Lab of Pattern Recognition, CASIA. His major research interests include machine learning, pattern recognition, and computer vision. He has widely published in highly ranked international journals such as IEEE TPAMI and IEEE TIP, and leading international conferences such as CVPR, ICCV, and ECCV. He has served as an Associate Editor of IEEE TPAMI, IEEE TIP, and PR. He is an IEEE Fellow and an IAPR Fellow.
\end{IEEEbiography}

\end{document}